\documentclass[referee]{aa}
\usepackage{graphicx}
\usepackage{rotating}

\def\spose#1{\hbox to 0pt{#1\hss}}                                            
\def\simlt{\mathrel{\spose{\lower 3pt\hbox{$\mathchar"218$}}
     \raise 2.0pt\hbox{$\mathchar"13C$}}}
\def\lsimeq{\mathrel{\spose{\lower 3pt\hbox{$\mathchar"218$}}
     \raise 2.0pt\hbox{$\mathchar"13C$}}}
\def\ls{\mathrel{\spose{\lower 3pt\hbox{$\mathchar"218$}}
     \raise 2.0pt\hbox{$\mathchar"13C$}}}
\def\simgt{\mathrel{\spose{\lower 3pt\hbox{$\mathchar"218$}}
     \raise 2.0pt\hbox{$\mathchar"13E$}}}                                     
\def\gsimeq{\mathrel{\spose{\lower 3pt\hbox{$\mathchar"218$}}
     \raise 2.0pt\hbox{$\mathchar"13E$}}}                                     
\def\gs{\mathrel{\spose{\lower 3pt\hbox{$\mathchar"218$}}
     \raise 2.0pt\hbox{$\mathchar"13E$}}}                                     
\def\cgs{erg cm$^{-2}$ s$^{-1}$}
\def\pn{\par\noindent}
\def\chandra{{\it Chandra}}
\def\xmm{XMM--{\it Newton}}
\def\cgs{erg cm$^{-2}$ s$^{-1}$}
\def\hel{{\tt Hellas2Xmm}}
\def\msun{M$_{\odot}$}

\begin{document}

\title{XMM--{\it Newton} observations of Extremely Red Objects 
and the link with luminous, X--ray obscured Quasars} 
\author{M. Brusa \inst{1,2,}$^{*}$,  A. Comastri\inst{2}, E. Daddi\inst{3}, 
L. Pozzetti\inst{2},
G. Zamorani\inst{2}, C. Vignali\inst{1,2}, A. Cimatti\inst{4},
F. Fiore\inst{5}, M. Mignoli\inst{2}, P. Ciliegi\inst{2},
H.J.A. R\"ottgering\inst{6}} 
\institute{Dipartimento di Astronomia Universita' di Bologna, 
    via Ranzani 1, I--40127 Bologna, Italy \\
    ($^{*}$ Present address: Max Planck Institut fuer Extraterrestrische Physik,
    D-85478 Garching, Germany. \\ E-mail: marcella@mpe.mpg.de) 
    \and INAF -- Osservatorio Astronomico di Bologna,  
    via Ranzani 1, I--40127 Bologna, Italy 
    \and European Southern Observatory, Karl--Schwarzschild--Strasse 2,
    D--85748 Garching bei Muenchen, Germany
    \and INAF -- Osservatorio Astrofisico di Arcetri, Largo E. Fermi 5, 
    I--50125 Firenze, Italy 
    \and INAF -- Osservatorio Astronomico di Roma, 
    via Frascati 33, I--00040 Monteporzio, Italy 
    \and Sterrewacht Leiden, PO Box 9513, 2300 RA, Leiden, The Netherlands }
\abstract{
We present the results of a deep (about 80 ks) XMM-Newton 
survey of the largest sample of near-infrared selected Extremely Red
Objects (R$-$K$>$ 5) available to date to K$_S\ls 19.2$.
At the relatively bright X-ray fluxes (F$_{2-10 \rm keV}\gs 4\times 10^{-15}$
\cgs) and near-infrared magnitude probed by the  
present observations, the fraction of AGN (i.e., X--ray detected) 
among the ERO population is small ($\sim$3.5\%); 
conversely, the fraction of EROs among 
hard X-ray selected sources is much higher ($\sim20$\%). 
The X-ray properties of EROs detected in our \xmm\
observation indicate absorption in excess of 10$^{22}$ cm$^{-2}$
in a large fraction of them. 
We have also
considered additional samples of X--ray detected EROs available in the
literature.  
X--ray spectral analysis of the highest S/N sources unambiguously indicates 
that large columns of cold gas (even $> 10^{23}$ cm$^{-2}$) are the
rule rather than the exception.
The X-ray, optical, and near-infrared properties of those X-ray
selected EROs with a spectroscopic or photometric redshift 
nicely match those expected for quasars 2, the
high-luminosity, high-redshift obscured AGNs predicted in baseline 
XRB synthesis models. 
A close correlation is detected between X- and K-band fluxes. For the
AGN EROs this is consistent, under reasonable assumptions, with the
relation established locally between the host galaxies and their central
black holes. This suggest that
the majority of EROs are powered by massive black holes accreting, on average, 
at about 0.03-0.1 of the Eddington limit.   
}
\authorrunning{M. Brusa et al.}
\titlerunning{XMM--{\it Newton} observations of EROs}
\maketitle
\section{Introduction}

Extremely Red Objects (EROs, R$-$K$>$5, Elston, Rieke \& Rieke 1988),
initially detected in near--infrared ground--based imaging, 
have the colors expected for high-redshift passive ellipticals  
and have been used as tracers of distant (z$\gs1$) and old spheroids.
Reproducing their observational properties have  
proved to be extremely challenging for all  current
galaxy formation models (see e.g. Kauffmann 2003 for a
review). 
However, on the basis of a number of observational results, it has
been pointed out that 
high-$redshift$ passive ellipticals are only one of the various classes 
of extragalactic sources which make up the ERO population.
Deep VLT spectroscopy from the {\tt $K20$ survey} (Cimatti et al. 
2002, 2003) has indeed shown that EROs are nearly equally populated 
by old, passively evolving systems and dusty star--forming galaxies 
over a similar range of redshift ($z=0.8-2$ for both the classes; see also
Yan, Thompson \& Soifer 2004) and similar results are confirmed both
by colour selection criteria (Mannucci et al. 2002) and by radio  
observations (Smail et al. 2002). 
A few individual objects have been also identified as 
high redshift Active Galactic Nuclei (AGN) on the basis of the
detection of strong emission lines in near--infrared and/or optical
spectra (see e.g. Pierre et al. 2001; Brusa et al. 2003). In this case
the enhanced emission in the K band with respect to the
R band is probably due to the combination of strong dust extinction
in the optical and a contribution of the point-like emission in the
near--infrared. However, there are increasing evidences
that the near--infrared light of {\it obscured} AGN is dominated by the host
galaxy emission (see e.g. Mainieri et al. 2002; Mignoli et al. 2004). \\
A large population of
optically faint X--ray sources without any obvious AGN signature in
the optical spectrum and with optical to near--infrared colors typical
of high redshift ellipticals and starburst galaxies has been revealed
in the deepest \chandra\ and \xmm\ exposures (e.g. Hasinger et al. 2001; Barger
et al. 2003; Szokoly et al. 2004). 
Thus, the follow--up campaigns of deep \xmm\ and \chandra\ observations
have probed to be a powerful tool to investigate the AGN EROs
population (Alexander et al. 2002).\\
Results from both shallow and deep X--ray surveys also suggested 
that the AGN population among EROs shares the same X--ray properties
of high--luminosity, highly obscured (N$_H>10^{22}$ cm$^{-2}$)
AGN (Mainieri et al., 2002; Alexander et al. 2002; Brusa
2003). 
Further support to the result that a significant fraction of 
obscured AGN are hosted in EROs comes from near infrared observations of
X--ray sources selected on the basis of their high X--ray to optical
flux ratio (X/O$>10$, Mignoli et al. 2004): the hosts of luminous, obscured
hard X-ray sources with extreme X/O are among the most
massive spheroids at z$\gs1$.
\pn
Finally, the observed fraction of AGN among EROs 
can help constraining models which include the evolution of QSO activity in
the formation of spheroids and the resulting effects on galaxy evolution
(e.g. Granato et al. 2004; Menci et al. 2004).
Several physical models have been proposed 
in which the fueling of the supermassive accreting black holes 
in AGN is triggered by merging events (in the context of the 
hierarchical structure formation paradigm), 
and the interplay between star formation and nuclear activity 
determines the relationship between the black holes (BH) mass 
and the mass of the host galaxy. 
If the evolution of luminous AGN follows that of spheroids, 
as suggested by e.g. Franceschini et al. (1999) and Granato et al. (2001), 
it is possible that the radiation and the strong 
winds produced by a powerful AGN present in the massive galaxy 
may help inhibiting the star-formation in these galaxies, which 
therefore would have red colors.
However, previous studies 
on the fraction of AGN among the EROs population,
although having deep near--infrared and X--ray observations ($K\sim
21-22$ and the Megaseconds {\it Chandra} exposures) 
were limited in areal coverage (50--80 arcmin$^{2}$) and therefore were
unsuitable for detailed statistical analyses of the AGN EROs
population.  \\
\pn
To further study the nature of AGN EROs and the link between accreting
supermassive black holes and the host galaxy properties,
we have started an extensive program of
multiwavelength observations of one of the largest sample of near--infrared 
selected EROs available to date ($\sim$400 sources), selected 
over a contiguous field of $\sim 700$ arcmin$^2$ (the ``Daddi field'',
Daddi et al. 2000). The sample is complete to a magnitude limit of
K$s\sim$19 and the field is covered by deep optical photometry in the
R--band. 
The same field will be also imaged with Subaru and \chandra\ and
spectroscopic VIMOS observations are already planned. 
We have obtained with \xmm\ a total of 110 ks, in two different
observations:
the moderate--deep exposure and the high energy throughput of
\xmm, coupled with its large field of view, are well--suited to detect
AGN among EROs at relatively bright X--ray fluxes, on a statistically
significant sample.  
The data reduction and analysis, the X--ray source identification 
and the X--ray properties of X--ray detected EROs in our \xmm\ sample
are presented in Section~2. The results on the fraction of AGN EROs
as a function of the X--ray and K--band fluxes are discussed
in Section~3. Section~4 compares the optical, near infrared
and X--ray properties of the EROs in our sample with those of 
other samples of X--ray detected EROs and discusses the fraction
of AGN EROs in K--selected samples.
Section~5 presents the average X--ray properties of EROs AGN,
their contribution to the quasar 2
population and an estimate of their Black Hole masses and
Eddington ratios. 
Finally, Section~6 summarizes the most important results.
Throughout the paper, a cosmology with $H_0=70$ km s$^{-1}$ Mpc$^{-1}$,
$\Omega_m$=0.3 and $\Omega_{\Lambda}$=0.7 is adopted. 

\section{Multiwavelength data and X--ray source identification}
\label{eros_id}
\subsection{Near--Infrared and Optical data}
\label{opt_data}
\pn
The near infrared EROs sample was selected by Daddi et
al. (2000) from the 5$\sigma$ K--band source catalog and 
adopting the selection criterion $R-Ks\geq 5$. 
The $R$-band data were taken at the 4.2m William Herschel Telescope
on La Palma, while the $Ks$ observations
were performed with the ESO NTT 3.5m telescope in La Silla; 
relevant details on optical and near--infrared data reduction 
can be found in Daddi et al. (2000). \\
A total of $\sim 400$ EROs are included in this ``reference EROs
sample''; to date, this study still constitutes the largest published 
survey of EROs performed at moderately deep $K$ limits, complete
to $Ks$=18.8 over $\sim 700$ arcmin$^2$ and to $Ks$=19.2 in a
deeper area of $\sim 450$ arcmin$^{2}$, more than a factor of  four
larger than other near--infrared surveys at the same limiting
magnitudes (e.g. Thompson et  al. 1999; Miyazaki et al. 2003). The
5$\sigma$ limiting magnitude in the $R$ band is $\sim25.5$.  

\subsection{X-ray data}
Two XMM--{\it Newton} observations of this field
have been obtained with the European Photon Imaging Camera (EPIC,
Jansen et al. 2001), equipped with both the  {\it MOS} and {\it pn}
instruments. 
The first observation (OBS-ID 0057560301) was taken on August 3, 2001 
for a nominal exposure time of 50 ks; the second 
observation was taken two years later, splitted in two parts (August
22, 2003 -- OBS-ID 0148520101 -- and September 16, 2003 -- OBS-ID 0148520301),
for a total nominal exposure time of $\sim 60$ ks. \\ 
All the EPIC cameras operated in full-frame and were equipped 
with the ``Thin'' filter, which is usually employed in the observations
of faint sources\footnote{http://xmm.vilspa.esa.es/external/xmm\_user\_support/documentation/index.shtml}. 
The three \xmm\ datasets were reduced using version 5.4.1 of
the Science Analysis
System\footnote{http://xmm.vilspa.esa.es/external/xmm\_sw\_cal/sas\_frame.shtml}
(SAS) with the latest, relevant calibration products.
The raw {\it pn} and {\it MOS} Observations Data Files (ODF) were processed
using the SAS tasks {\tt emproc} and {\tt epproc} to produce calibrated 
event lists.
Only events with pattern 0-4 (single and double) for the {\it pn} and 0-12
for the MOS cameras were selected. 
All the event files were cleaned up from hot pixels and soft
proton flares
removing all the time intervals with a count rate
higher than 0.15 c/s in the 10--12.4 keV energy range for the MOS and
higher than  0.35 c/s in the 10--13 keV band for the {\it pn} units
(see Baldi et al. 2002).
\pn
The excellent {\it relative} astrometry between the three cameras in each
observation (within 1$''$, well below their PSF FWHM of $\sim6''$), allowed us
to merge the \textit{MOS} and \textit{pn} images in each observation,
thus increasing the signal-to-noise ratio and reaching fainter
X--ray fluxes. Moreover, taking into account the {\it absolute}
astrometry between the three observations, the counting statistics 
have been improved by summing all the available datasets (i.e. {\it
MOS1}, {\it MOS2} and {\it pn} of the three observations). 
The resulting total exposure time for the {\it pn} is $\sim 82$ ks,
and is only slightly lower for the {\it MOS} instruments ($\sim 78$ ks).\\
\pn
\begin{figure}[!t]
\center
\includegraphics[width=0.8\textwidth]{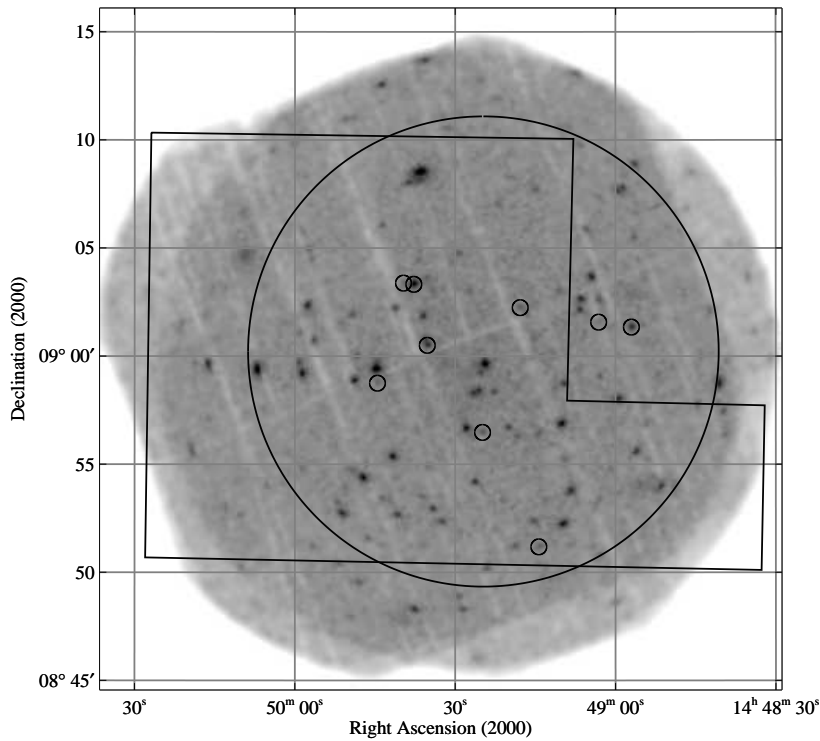} 
\caption{X--ray image in the 0.5-10 keV band (grey scale) with
superimposed the 11 arcmin radius circle of the X--ray analysis and
the deepest region ($\sim 450$ arcmin$^{2}$) in the K-band (polygon).
The shallower K--band image covers the entire X--ray field of view.
The X--ray sources associated with EROs are marked with empty circles.
}
\label{xmm_nir}
\end{figure}
We have limited the X--ray analysis to a circular region of 11
arcmin radius from the point of maximum exposure time (centered at 
$\alpha$=14$^{h}$49$^{m}$25$^{s}$ and
$\delta$=09$^{o}$00$^{'}$13$^{''}$). At this radius,  
the effective exposure drops to $\sim 50$\% of the maximum value.
The resulting area analysed in the present work is therefore of $\sim 380$
arcmin$^2$. 
The area in common with the deepest near--infrared coverage is $\sim
300$ arcmin$^2$.
The pattern of the X--ray and optical/near--infrared data is shown in
Fig.~\ref{xmm_nir}: the 11 arcmin radius circle and 
the deepest K$_s$ band region (polygon) are superimposed on the
0.5-10 keV image.\\
The \xmm\ area analysed in this work includes 257 EROs:
173 EROs with $Ks\ls$18.8 over $\sim 380$ arcmin$^{2}$, and 
216 EROs with $Ks\ls$19.2 in the area with the deeper
near--infrared coverage ($\sim 300$ arcmin$^2$).
The accurate detection algorithm developed for the \hel\ survey
(see Baldi et al. 2002 for details) was run on the 0.5--10 keV (full
band) cleaned events, in order to create a preliminary list of
candidate sources, which included also sources detected at a low level
of significance.  
We then computed for each source the probability that the 
detected counts originate from poissonian background fluctuations
and we kept in the final list only those sources (96) which 
were detected above a detection probability threshold of
{\it p}=2$\times 10^{-5}$ (that corresponds to less than 1 spurious
X--ray detection, or $\simeq$ 4$\sigma$ gaussian threshold).
The count rate to flux conversion factor was derived assuming a
power law with photon index $\Gamma$=1.7, 
absorbed by the Galactic column density in the direction of the Daddi Field
(N$_H$=5$\times 10^{20}$ cm$^{-2}$, Dickey \& Lockman 1990), 
and weighted by the effective exposure times of
the different EPIC cameras. 
The uncertainty in the derived fluxes is $<15$\% for 
$\Delta \Gamma=\pm0.5$ and N$_H$ up to 10$^{21}$ cm$^{-2}$.
The corresponding full band limiting flux is 
$\sim 2 \times 10^{-15}$ \cgs at
the aim point, and is a factor of $\sim 1.5$ higher at the 
edge of the selected area, where the net exposure is $\sim 40$ ks. \\
In order to compute reliable X--ray fluxes in different energy bands
(i.e. to roughly account for the X--ray spectral shape), 
fluxes in the soft (0.5--2 keV band) and hard (2--10 keV band)
were computed from the counts detected in each energy band
using the same detection algorithm. A total of 86 sources (down to
S$_{lim}\sim9\times 10^{-16}$ \cgs) and 60 sources
(S$_{lim}\sim4\times 10^{-15}$ \cgs) were detected, respectively.
\pn
The X--ray centroids have been astrometrically calibrated 
with respect to the optical positions of three bright quasars 
in the field (Hall et al. 2000): the resulting shift of $\sim$ 2$''$ 
($\Delta$(Ra)=1.47$''$; $\Delta$(dec)=$-$1.29$''$) has then been
applied to all of the source positions.\\
Table~\ref{tab+xray} lists all the relevant X--ray properties of the detected
sources (the X--ray source identifier, X--ray coordinates,
hard 2-10 keV flux and counts, soft 0.5--2 keV flux and counts),
sorted with decreasing hard band X--ray flux.  
%
\subsection{Likelihood analysis}
\begin{figure}[!t]
\includegraphics[width=0.9\textwidth]{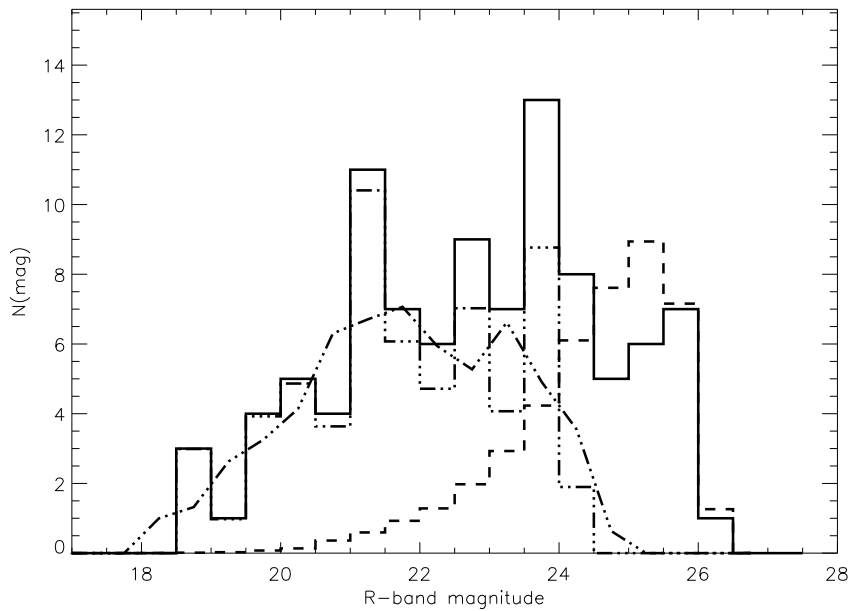} 
\caption{
Observed magnitude distribution of the
optical objects detected in the $R$ band within a radius of 3\arcsec\
around each of the 87 X--ray sources (solid histogram), together with the
expected distribution of background objects in the same area ({\it
n(m)}, dashed histogram). The difference between these two
distributions (dot-dot-dot dashed histogram) is the expected magnitude
distribution of the optical counterparts. The smooth curve fitted to
this histogram  (dot-dot-dot dashed line) has been used as input in
the likelihood calculation ({\it q(m)}).} 
\label{histo}
\end{figure}

\pn
Optical and near--infrared photometry is available for 87 (55) out of
96 (60) X--ray sources detected in the full (hard) band. 
The sources for which optical and near--infrared photometry is not
available are all close to bright stars and/or defects in the $R$ and/or $Ks$
band images that were masked in the optical and near--infrared source
detection.
\pn
At the optical and near--infrared fluxes probed by our survey,
the identification process is a critical issue especially 
for faint sources. 
At first, the X--ray source list has been cross--correlated with the 
$K$--band and $R$--band catalogs using a conservative 5$''$ radius 
error circle (see Brusa et al. 2003 for further details): 162 (71) sources
in the $R$ ($Ks$) band were found in 87 X--ray error-circles. 
The difference in the number of objects (i.e. surface densities) 
in the two bands clearly reflects the different depth of the optical
and near--infrared images. 
In particular, 
using the $R$ catalog and a fixed searching radius, substantial
source confusion may be present:
on the basis of the integral counts from the $R$--band catalog, on average 
1 galaxy with R$<25$ is expected just by chance 
in each of the 5$''$ radius error circles. 
We therefore decided to use the ``likelihood ratio'' ($LR$) technique,
in order to properly identify the optical/near--infrared counterparts.
The $LR$ is  
defined as the ratio between the probability that the source is
the correct identification and the corresponding probability of being a
background, unrelated object (Sutherland \& Saunders 1992), i.e.: 
\begin{equation}
LR = \frac{q(m) f(r)}{n(m)}
\end{equation}
where {\it f(r)} is the probability distribution function of the
positional errors and it is assumed to be a two--dimensional gaussian,
{\it n(m)} is the surface density of background objects with magnitude
{\it m}, and {\it q(m)} is the expected probability distribution as a
function of magnitude of the true counterparts.
The {\it q(m)} distribution is normalized as $\int^{m_{faint}}q(m)dm =
Q$, where $Q$ is the {\it a priori} expected fraction 
of X--ray sources with an optical counterpart brighter than
$m_{faint}$; $m_{faint}$ can be either the limiting magnitude of the
optical data or the magnitude beyond which the surface density of
background objects becomes so high that no reliable ``statistical''
identification is possible. \\ 
For the calculation of the $LR$ parameters 
we have followed the procedure described by Ciliegi et
al. (2003); more specifically, 
in order to maximize the statistical significance of
the over density due to the presence of the optical counterparts, we
have adopted a 3\arcsec\ radius for the estimate of the {\it q(m)} 
distribution.
A large fraction of the possible counterparts are expected to be included 
within such radius, on the basis of previous works on \xmm\ data
(e.g. Fiore et al. 2003).\\
Fig.~\ref{histo} shows the observed magnitude 
distribution of the optical objects detected in the $R$ band 
within a radius of 3\arcsec\
around each X--ray source (solid histogram), together with the
expected distribution of background objects in the same area (dashed
histogram). The difference between these two distributions
(dot-dot-dot dashed histogram) is the expected magnitude distribution
of the optical counterparts. The smooth curve fitted to this histogram
(dot-dot-dot dashed line) has been used as input in the likelihood
calculation ({\it q(m))}.  
Figure~\ref{histo} shows that the observed number of 
objects is well above the  background up to $R\sim24$. At $R>24$, the number of
detected objects in the X--ray error boxes is consistent with 
that expected from the background. 
For these reasons we have adopted $R_{faint}\sim 24$ in our likelihood
calculation; all sources fainter than this limit will have {\it
q(m)=0}  by definition and correspondingly $LR=0$. 
For the $Q$ normalization we adopted $Q=0.75$, corresponding to the ratio
between the integral of the $q(m))$ distribution and the total
number of X--ray sources. This preliminary analysis suggests that  
we expect to identify a fraction of the order of 75\% of the X--ray 
sources with objects brighter than R$\sim 24$,
in agreement with the results from other \xmm\ surveys (e.g. Hasinger
et al. 2001).
A similar procedure has been applied to the $K$ band data and the $LR$
value for all the optical and near--infrared candidates  
has been computed. \\ 
The next step is to choose
the best threshold value for $LR$ ($L_{\rm th}$) to
discriminate between spurious and real identifications. 
The choice of $L_{\rm th}$ depends on two factors: first, it
should be small enough to avoid missing many real
identifications and producing a rather incomplete sample.
Secondly, $L_{\rm th}$ should
be large enough to keep the number of spurious identifications
as low as possible and to increase the reliability of the source
identifications.  
A $LR$  threshold of $L_{\rm th}$=0.25 in both optical and
near--infrared bands has been adopted; this turned out to be the value
which maximizes the sum of sample reliability and completeness
for the assumed $Q$ normalization (see Ciliegi et al. 2003 for further
details). To check how the uncertainty in Q could affect our
results, we repeated the likelihood ratio analysis using different
values of $Q$ in the range 0.5--1.0: no substantial difference in the
final number of identifications has been found. \\
This threshold, {\it 
a posteriori}, led to an estimated percentage of secure X--ray to
optical or near--infrared associations up to $m_{faint}$ of the order
of $\sim 80$\%, in good agreement with the estimate of Q. \\ 

\subsection{X--ray source identification}
The information derived in the two bands have been then combined:
all the sources with the highest $LR>L_{\rm th}$ in both the $R$ and 
$K$ bands, as well as the sources undetected in the $K$ band but with
a $LR>L_{\rm th}$ in the $R$--band have been defined
{\it secure} identifications (a total of 70).
As expected, most of the reliable optical counterparts have an X--ray to
optical separation ($\Delta(X-O)$) smaller than $3''$, with only 
4 objects with $3.16''< \Delta(X-O)<3.71''$. \\
In addition, three X--ray sources have a 
unique, very faint (24.5$\ls R \ls 25.2$) optical counterpart  
within 1.6$''$ from the X--ray position 
(xid\#330, xid\#244, xid\#129)\footnote{These sources 
have no further optical counterpart up to 5$''$}; by construction 
(see Sect. 2.3), 
their associated likelihood ratio is zero. Given that {\it less than one}
galaxy with $24 \ls R \ls 25$ is expected by chance  
in the total area corresponding to 87 error boxes with
1.6$''$ error--box, 
we tentatively consider also these three sources as likely identifications.
Seventy--three out of 87 X--ray sources have been 
therefore securely associated to an optical/near--infrared 
counterpart and are reported in the first part of
Table~\ref{tab_xray_opt},
where we list, for each X--ray source, 
the X--ray ID number and position, the distance between the X--ray and
optical counterparts (or candidates), the $R$ and $K$--band
magnitudes, and the $LR$ in both the optical and near--infrared bands.  
Of these 73 sources, 45 belong to the hard X--ray sample. 
In other 8 cases we have more than one optical and/or near--infrared 
possible counterpart with $LR>LR_{\rm th}$ 
and therefore the real counterparts are not unambiguously
determined. These are
listed in the second part of Table~\ref{tab_xray_opt}.  
In these cases, a detection in the forthcoming \chandra\ observations 
(with 5-10 times smaller
error--boxes) would definitively discriminate between the possible
counterparts.\\ 
Finally, in the remaining 6 cases the possible
counterparts have on average fainter optical magnitude, 
none of the candidates has $LR>$ $L_{\rm th}$ , and all of them lie 
at $\Delta(X-O)>3''$ (Table~\ref{tab_xray_opt}). 
This can be the case if the source is very faint and undetected 
in the optical bands (see e.g. Koekemoer et al. 2004), 
or if the X--ray emission originates from a group of galaxies.
In the bottom part of  Table~\ref{tab_xray_opt} we list also the 9 X--ray 
sources for which optical and infrared photometry is not available.
In the following, we will consider only the ERO sources;
a more detailed discussion of the properties of the
global sample of optical/near--infrared identifications will be
presented elsewhere (Brusa et al., in preparation).

\subsection{Hard X--ray detected EROs and hardness ratio analysis}
From the likelihood analysis, 8 hard X--ray sources are
securely associated with  EROs in the ``reference EROs sample''
(i.e., in the 5$\sigma$ catalog, see Sect.~\ref{opt_data}). 
All the EROs associated  with X--ray sources are reported in
Table~\ref{tab_eros_1}. 
Differently from Table~\ref{tab_xray_opt}, the $R$ and $K$ magnitudes 
given here are measured within 2$''$ diameter aperture. These are the
magnitudes used to compute the $R-K$ colour and to select the EROs
sample.
The same table gives, for each source, the hardness ratio (HR) 
defined as (H-S)/(H+S) where H and S are the counts in 
the hard and soft band, respectively. Among the 8 X--ray detected 
EROs, 5 are detected in both the
hard and soft bands and three only in the hard (lower limit to
HR). \\
One more X--ray source is associated with an ERO with a $K$
magnitude fainter than the 5$\sigma$ threshold (bottom part of
Table~\ref{tab_eros_1}).
A total of 9 hard X--ray sources are therefore associated with EROs. \\
\begin{figure}[!t]
\center
\includegraphics[width=0.8\textwidth]{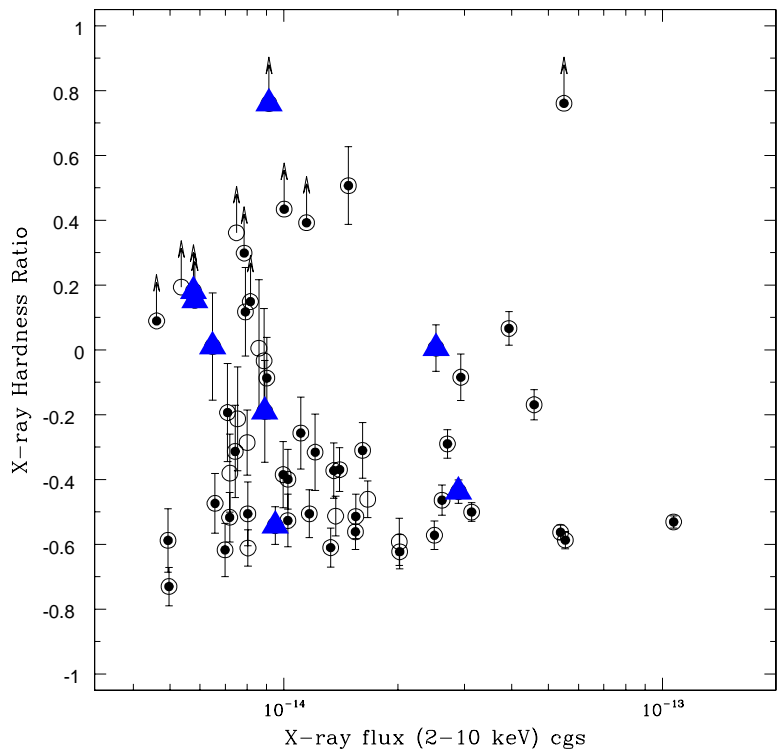}  
\caption{The hardness ratio versus 2--10 keV flux 
for the hard X--ray selected sources in the ``Daddi Field''. 
 Sources with a secure optical ID (i.e. first part of Table 1) 
are indicated as filled symbols. The 8 hard X--ray detected EROs
from the 5 $\sigma$ sample are plotted as filled triangles (see Table 3)}.
\label{hr_flux}
\end{figure}
Figure~\ref{hr_flux} shows the HR versus the 2--10 keV flux 
for the 60 hard X--ray selected sources. The eight EROs
belonging to the 5 sigma sample are plotted as filled triangles.
Six out of eight EROs have HR values higher than the median
value of the total sample (HR$_{med}$=-0.3); the HR of these 6 EROs
imply N$_H>10^{22}$ cm$^{-2}$ at z$>1$ (see Fig.~7 in Brusa et
al. 2003).
Only two EROs have HR values in the range expected for 
unobscured  AGN (HR$\sim-0.4$), and they are 
detected at 
F$_{2-10 keV}\gs 10^{-14}$ \cgs, while the majority 
of the EROs are detected at 
F$_{2-10 keV}\ls 10^{-14}$ \cgs and hard X--ray colours, 
qualitatively in agreement with the observed  hardening of the 
X--ray source population at fainter flux levels (e.g., Tozzi et al. 2001).\\
These data strongly suggest that significant absorption 
is present in a large fraction of the hard X--ray detected EROs
population (see also Sect. 5). 
On the other hand, Figure 3 shows that even if the fraction of EROs
increases among sources with higher values of HR, not all of the
hardest sources are EROs (e.g. among the 15 sources with HR$>$0, only
6 are EROs).

\section{Fractions of X--ray detected EROs and of EROs among hard
X--ray sources}
\label{frac_eros}
The large area of our sample is well suited
to statistically assess the fraction of
AGN powered EROs at relatively bright X--ray fluxes and to
quantitatively investigate the luminous tail of this population. \\
257 K--selected EROs from the ``reference EROs sample''
are within the \xmm\ area analysed in this work.
Among the 9 hard X--ray detected objects, 
seven are in the deeper ($Ks\ls 19.2$) 300 arcmin$^{2}$
(see Fig.~\ref{xmm_nir}),  
while the number of X--ray detected EROs 
in the 380 arcmin$^{2}$ area at the shallower K$s<18.8$ 
limit is 6; one additional object
(xid\#189) is detected over the incomplete 18.8$\ls Ks\ls 19.2$ area 
(see also Table~\ref{tab_frac}).\\
In the present observation, the X--ray limiting flux corresponds to an
X--ray luminosity L$_X\gs 10^{43}$ erg s$^{-1}$ for z=1. 
Thus, the EROs X--ray emission is most likely 
powered by AGN activity, and the fraction of
AGN among EROs in the present sample is {\it at least}
$3.2\pm1.7$\% (8/257).  
The corresponding 
surface densities of X--ray emitting EROs are $\sim
1.6\pm0.6\times10^{-2}$ arcmin$^{-2}$ (6 over 380 arcmin$^{2}$) and
$2.3\pm0.9\times10^{-2}$ arcmin$^{-2}$ (7 over 300 arcmin$^{2}$) at
$Ks\ls 18.8$ and $Ks\ls 19.2$, respectively.   \\
The fraction of X--ray detected EROs in K--selected samples
has been reported by Alexander et al. (2002; hereafter A02) and
Roche, Almaini \& Dunlop (2003; hereafter R03). 
A02 detected 6/29 EROs\footnote{We note that
these authors adopted an $I-K>4$ selection for the definition
of their EROs sample, that roughly corresponds to a $R-K>5.3$
selection.}  (21\%) in the 
{\it Chandra} Deep Field--North (CDF--N) observation, 
where the $K$--band limit is about one
magnitude fainter ($Ks=20.1$) and the X--ray limiting flux is about
one order of magnitude deeper than in our observation, with a
corresponding limiting luminosity of L$_X\simeq 10^{42}$ \cgs\ at z=1.
At the faint X--ray fluxes probed by ultra--deep \chandra\ exposures, 
starbursts and normal elliptical galaxies start to be detected 
and they are usually characterized by softer X--ray
colors; a conservative estimate of the
fraction of AGN powered EROs in the CDF--N based on the hard X--ray
detections and luminosities (see also Sect.~\ref{erosx}) is  
14$^{+11}_{-7}$\% (A02; Vignali et al. 2002). \\
\scriptsize
\setcounter{table}{3} 
\begin{table*}[!t]
       \caption{Fraction of Hard X--ray detected EROs}
\label{tab_frac}
\vskip 4pt
\begin{tabular}{lccccc}
\hline
\hline
K(lim) & F$_{2-10 keV}$(lim) & area & N(EROs) & N(EROs) & \% \\
  & \cgs\ & arcmin$^{2}$ & K--selected  & X--ray detected & \\
\hline
This work: & & & & \\
18.8 & 4$\times10^{-15}$ & $\sim$380 & 173 & 6 & 3.5\% \\
19.2 & 4$\times10^{-15}$ & $\sim$300 & 216 & 7 & 3.2\% \\
all sample & 4$\times10^{-15}$ & $\sim$380 & 257 & 9 & 3.5$\pm1.2$\% \\
\hline
A02$^{\dag}$: & & & & & \\
20.1 & 2$\times10^{-16}$ & $\sim$70 & 29$^\ddag$ & 4 & 14$^{+11}_{-7}$\% \\
\hline
R03$^{\dag}$: & & & & \\
21.5 & 4$\times10^{-16}$ & $\sim$50 & 179 & 12 & 6.6$^{2.0}_{-1.8}$\% \\
\hline
\hline
\end{tabular}
\vskip 2pt
{\baselineskip 9pt
\footnotesize
\indent
$^{\dag}$: AO2: Alexander et al. 2002; R03: Roche, Almaini \& Dunlop
2003. \\
$^{\ddag}$: selected on the basis of a $I-K>4$ criterion.\\
}
\vglue0.2cm
\end{table*}
\normalsize
\setcounter{table}{0} 
From the R03 sample it is possible to estimate the fraction of
hard X--ray detected EROs in the CDF--S GOODS area, at an X--ray limiting flux
comparable to that of the CDF--N sample but extending down to significantly
fainter near--infrared magnitudes ($Ks\sim22$): about 6.6\% of the
$Ks$-selected EROs are associated with hard X--ray sources. \\ 
Even if the estimates from A02 and R03
at the faintest K magnitudes are obtained over
very small areas and may suffer from substantial cosmic variance, 
the differences in the fractions of X--ray detected EROs observed 
in these three samples are likely to be mainly due to the
combination of different X--ray and near--infrared limiting fluxes, as 
discussed in Sect.\ref{frac_k_sect}.\\
\pn
With \xmm\ and \chandra\ surveys, the fraction of optical
counterparts with extremely red colors has significantly increased
with respect to the first examples of EROs found in deep ROSAT
observations in the Lockman Hole (Lehmann et al. 2001).
The present data imply that a fraction of the order of about
16--18\% of the hard X--ray selected \xmm\ sources exhibit R$-$K$>5$ colors
(9/55 or 9/49 considering only the secure X--ray to optical associations).
Our results are in agreement with those reported by Mainieri et al. (2002)
in the Lockman Hole:
12/53 ($\sim$23\%) of hard X--ray selected sources are associated with EROs, 
at limiting near--infrared and X--ray fluxes comparable with those 
of the present sample. 
This fraction is about the same in the R03 sample 
($\sim 23\%$) and in the CDF--N sample ($\sim 21$\%, from the Barger et
al. 2003 catalog).  \\
\section{Multiwavelength properties of AGN EROs}
\subsection{The ``literature sample'' of hard X--ray detected EROs}
\label{comparison}
In order to investigate the nature of hard X--ray selected EROs
and the link between faint hard X--ray sources and the 
ERO population, we have collected all the literature multiwavelength data
available to date for EROs individually detected in the hard (2--10
keV) X--rays and selected on the basis of a R$-$K$>5$ criterion.
More specifically: 
\begin{itemize}
\item[$\bullet$] 1) Seventy EROs detected in the 2--8 keV band in the
CDF--N observation, from the Barger et al. (2003)
catalog; 9 have spectroscopic 
redshifts and 25 have photometric redshift estimates;
\item[$\bullet$] 2) Twenty--two hard X--ray detected EROs in the
\chandra\ Deep Field--South (CDF--S) observation, from Szokoly 
et al. (2004); 8 have spectroscopic redshift;   
\item[$\bullet$] 3) Twelve EROs detected in the hard (2--10 keV) band in
the \xmm\ Lockman Hole observation (Mainieri et al. 2002); 
2 have spectroscopic redshifts and 3 have photometric redshift;
\item[$\bullet$] 4) Ten hard X-ray detected EROs from the \hel\
survey selected on the basis of an X--ray to optical flux ratio 
X/O$>10$ and R$>24$; all of them with redshifts estimated on the basis
of the observed R$-$K colors (Mignoli et al. 2004);  
\item[$\bullet$] 5) Five additional hard X-ray detected EROs available
in the literature (``additional sample'': Gandhi et al. 2004; Crawford
et al. 2002; Brusa et al. 2003; Willott et al. 2003); 4 have
spectroscopic redshifts and 1 photometric redshift.               
\end{itemize}
This literature sample consists of 128 EROs, 
including the 9 EROs discussed in the present work, detected in the
2--10 keV band; for 62 of them  photometric or spectroscopic redshifts 
are available.
This sample is by no means homogeneous and complete, 
but can be considered representative of EROs individually 
detected in the X--rays. \\
\pn
\begin{figure}[!t]
\center
\includegraphics[width=0.48\textwidth]{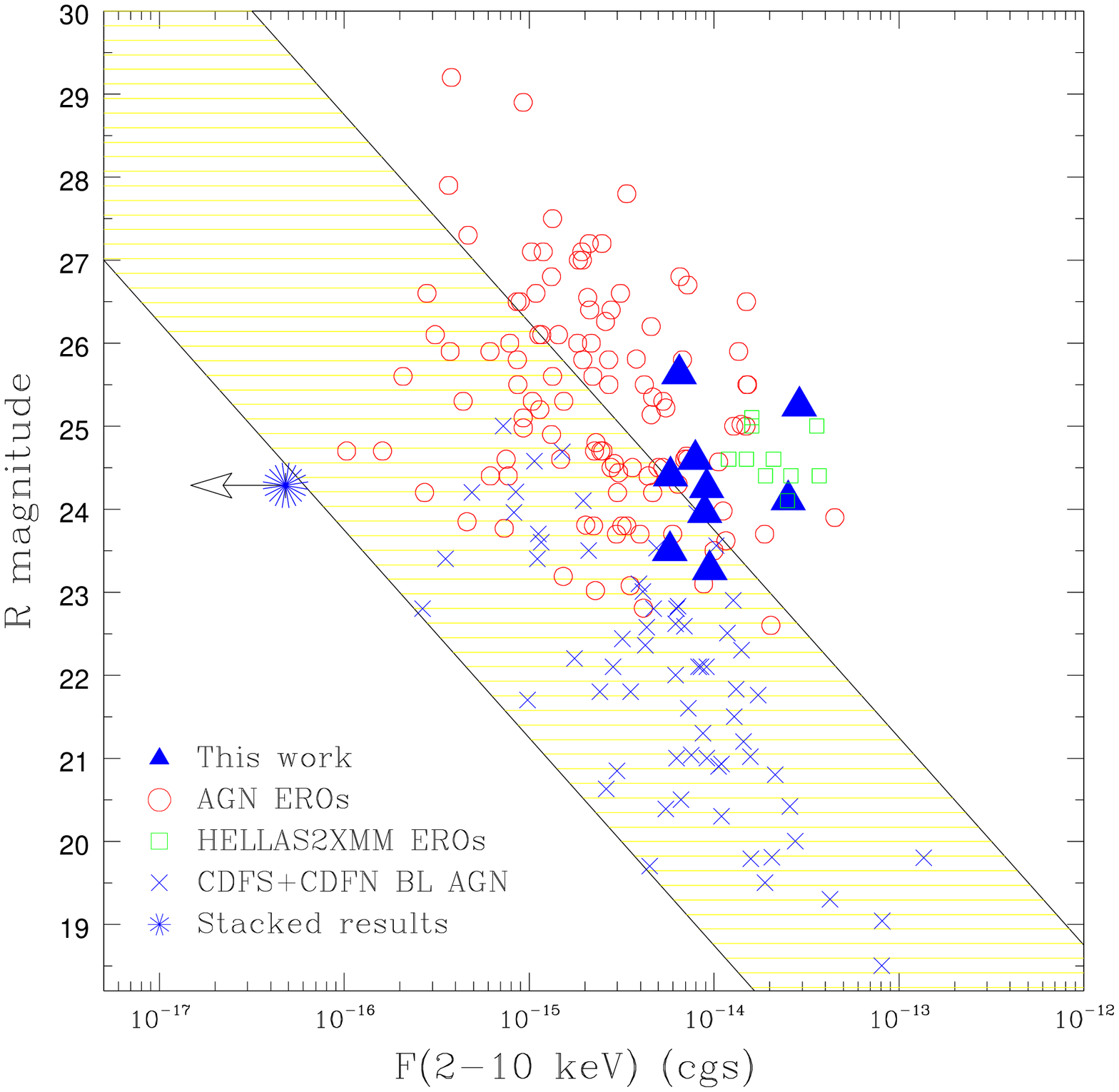}  
\includegraphics[width=0.48\textwidth]{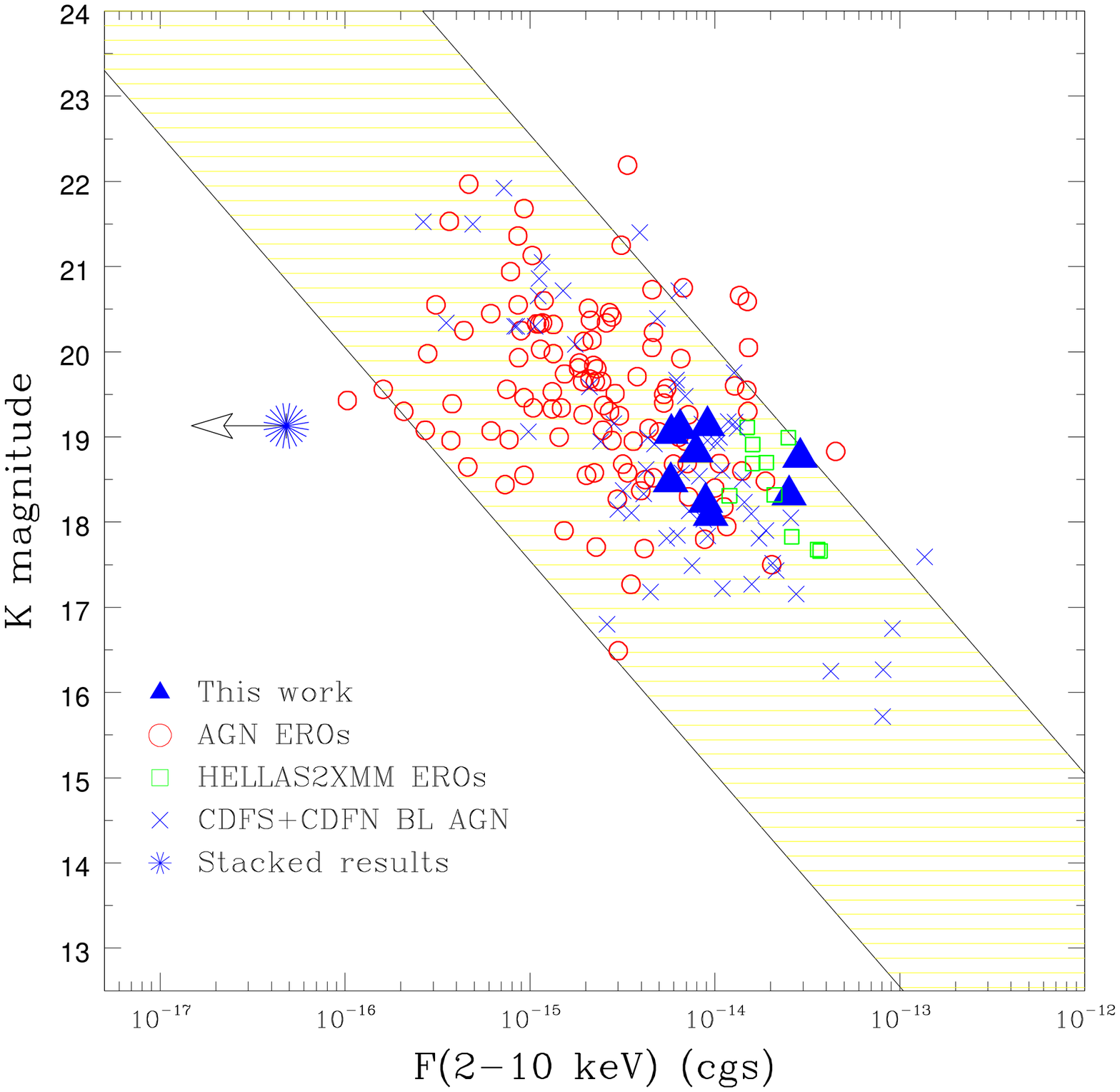}  
\caption{{\it left panel:} R--band  magnitude vs. hard X--ray flux
for EROs, serendipitously detected in hard X--ray surveys. 
Large filled triangles are the 9 hard X--ray selected EROs of this work;
circles correspond to the EROs in the ``literature sample'';
squares are sources from the \hel\ survey.
As a comparison,  broad--line AGN, i.e. sources with broad optical
emission lines in the optical spectrum, detected in the CDF--S and 
CDF--N surveys are also reported as crosses.
The shaded area represents the region typically occupied by known AGN 
(e.g. quasars and Seyferts) along the correlation
log$(X/O)=0\pm 1$.
For comparison, we report the result of the stacking  
analysis performed on the K20 EROs in the 
CDF--S field not individually detected in the \chandra\ observation 
(asterisk at the faintest X--ray flux; Brusa et al. 2002).
{\it Right panel:} the same plot but in the
K--band.}
\label{rxeros}
\end{figure}
\subsection{X--ray to optical/near--infrared properties of AGN EROs}
\label{eros_xo}
This enlarged ERO sample spans a wide range of optical and
hard X--ray fluxes. 
The R--band magnitudes plotted versus the hard X--ray
fluxes for all the 128 EROs in the literature sample are reported 
in Fig.~\ref{rxeros} (left panel): the 9 EROs from this work are
reported as triangles, the 10 EROs from the \hel\ survey
as squares and the remaining objects as circles.
In the same figure, the sources 
classified as broad line (BL) AGN in the CDF--N and CDF--S optical
catalogs (Barger et al. 2003; Szokoly et al. 2004) are also reported
as crosses. \\ 
X--ray detected EROs show an average
ratio between the X--ray and optical fluxes\footnote{The R--band flux is computed by
converting R magnitudes into monochromatic  fluxes and then multiplying 
them by the width of the R filter (Zombeck 1990). 
For a given X--ray energy 
range and R--band magnitude the following relation holds: 
$log ({\rm X/O}) = log {\rm f_X} + {\rm R}/2.5 + const$
where f$_{\rm X}$ is the X--ray flux, R is the optical magnitude and 
{\it const} depends only on the R--band filter used
in the optical observations; 
an indicative, average value is {\it const}=5.5 
(see Hornschemeier et al. 2000) and it can be used when datasets from
different observations are compared.} (X/O)  
around X/O$\simeq$10, about one order of magnitude higher than that
found for BL quasars by {\tt ROSAT} (Hasinger et al. 1998;
Lehmann et al. 2001) and recently extended by {\it Chandra} and
XMM--{\it Newton} observations 
(Alexander et al. 2001; Rosati et al. 2002; crosses in
Fig.~\ref{rxeros}). We note that  
the same shift with respect to the majority of quasar population is observed 
also excluding  the 10 \hel\ sources selected on the basis of 
their high X/O ($>10$). \\
The observed X--ray to optical properties of X--ray detected EROs 
are different also from that of the majority of near--infrared
selected EROs: 
the results of the stacking analysis of EROs 
not individually detected in the X--rays in the {\tt K20}
survey (asterisk in Fig.~\ref{rxeros})
led to an average X/O which is at least two order of magnitudes
lower than that of the EROs in the present sample
(Brusa et al. 2002; see also Alexander et al. 2002). \\ 
Obscured accretion at high redshifts is the most likely 
mechanism for explaining 
the observed X--ray to optical properties.
Moving the Spectral Energy Distribution (SED) of an X--ray absorbed
AGN to progressively higher redshifts the K--corrections 
in the optical and X--ray band work in the opposite direction. 
The ratio between the optical to X-ray optical depth, in the observer frame,
scales roughly as $(1+z)^{3.6}$, because dust extinction increases in
the UV while X-ray absorption strongly decreases going toward high
energies.  The net result is that in the presence of an absorbing
screen the observed optical flux of a high-z AGN can be strongly
reduced, and the observed magnitudes are mainly due to starlight in
the host galaxies. 
Conversely, the 2-10 keV X-ray flux can be much
less reduced.  Many extreme X-ray to optical ratio sources could then
be highly obscured quasar, i.e. type 2 QSO
(Fiore et al. 2003; Comastri, Brusa \& Mignoli 2003). 
The observed high values of the X/O are therefore at least
qualitatively consistent with those expected by a 
population of high redshift, absorbed AGN with X--ray column densities
in the range N$_{H}$=10$^{22}$-$10^{24}$ cm$^{-2}$
(see also Sect.~\ref{erosx}). \\
\pn
The right panel of Fig.~\ref{rxeros} shows that in the 
K--band magnitude vs. X--ray flux plane the X--ray detected EROs
occupy essentially the same region as the broad-line AGN,
with an average X--ray to near infrared ratio (X/K) of the order of 1.
Given that the K--band is less affected by absorption, the
fact that AGN EROs are indistinguishable from the overall quasar
population in this plane supports the hypothesis that their high
X/O ratios are mainly due to significant {\it nuclear} 
extinction in the optical band (see also Mainieri et al. 2002).
Moreover, the observed X/K correlation implies that, for a given  
hard X--ray flux, the $K$ magnitude can 
be predicted reasonably well ($\sigma\sim1.3$ mag) with a single 
relation both for broad line and ERO AGN. \\
\begin{figure}[!t]
\center
\includegraphics[width=0.8\textwidth]{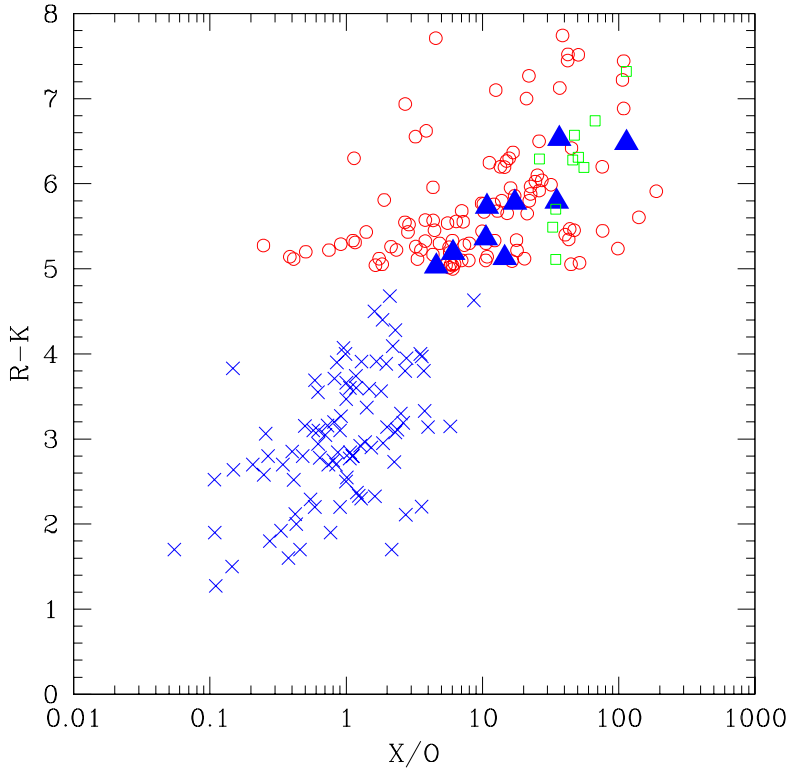}  
\caption{The R--K color as a function of the X--ray to optical flux
ratio (X/O) for EROS in the literature sample (empty circles are
objects from the deep fields; filled triangles the EROs from this work;
empty squares sources from Mignoli et al. 2004) and for BL AGN
selected in the CDF-N and CDF-S fields (crosses). 
See text for details.}
\label{rk_xo}
\end{figure}
The relationship between the hard X--ray and near--infrared fluxes
can also explain the trend observed between the R--K 
colors and the X/O for all the EROs 
in the literature sample compared with that observed for BL AGN
(Fig.~\ref{rk_xo}): the higher the X/O is, the redder the source is. \\

\subsection{AGN in K--selected EROs samples}
\label{frac_k_sect}
\begin{figure}[!t]
\center
\includegraphics[width=0.48\textwidth]{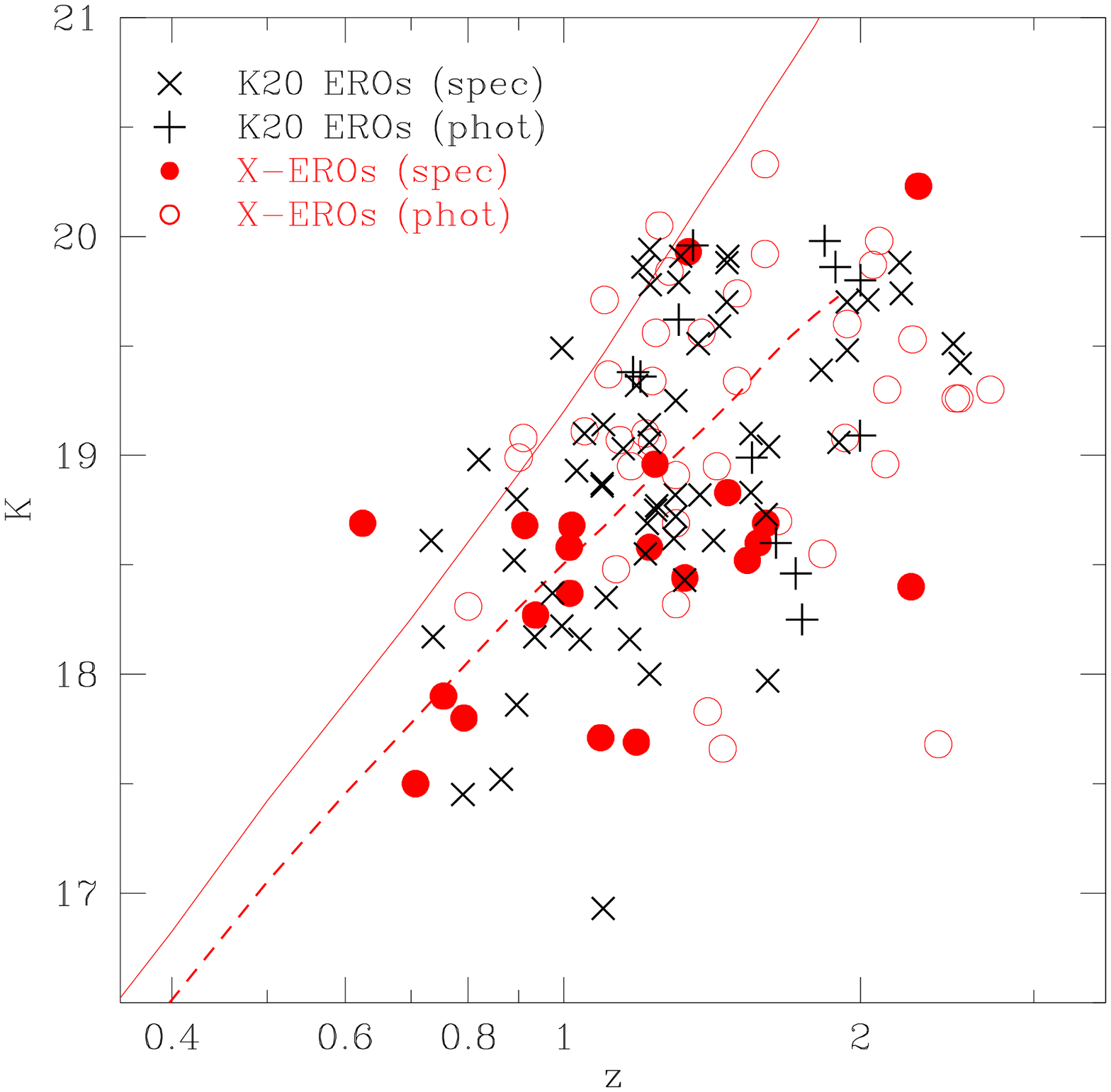}  
\includegraphics[width=0.48\textwidth]{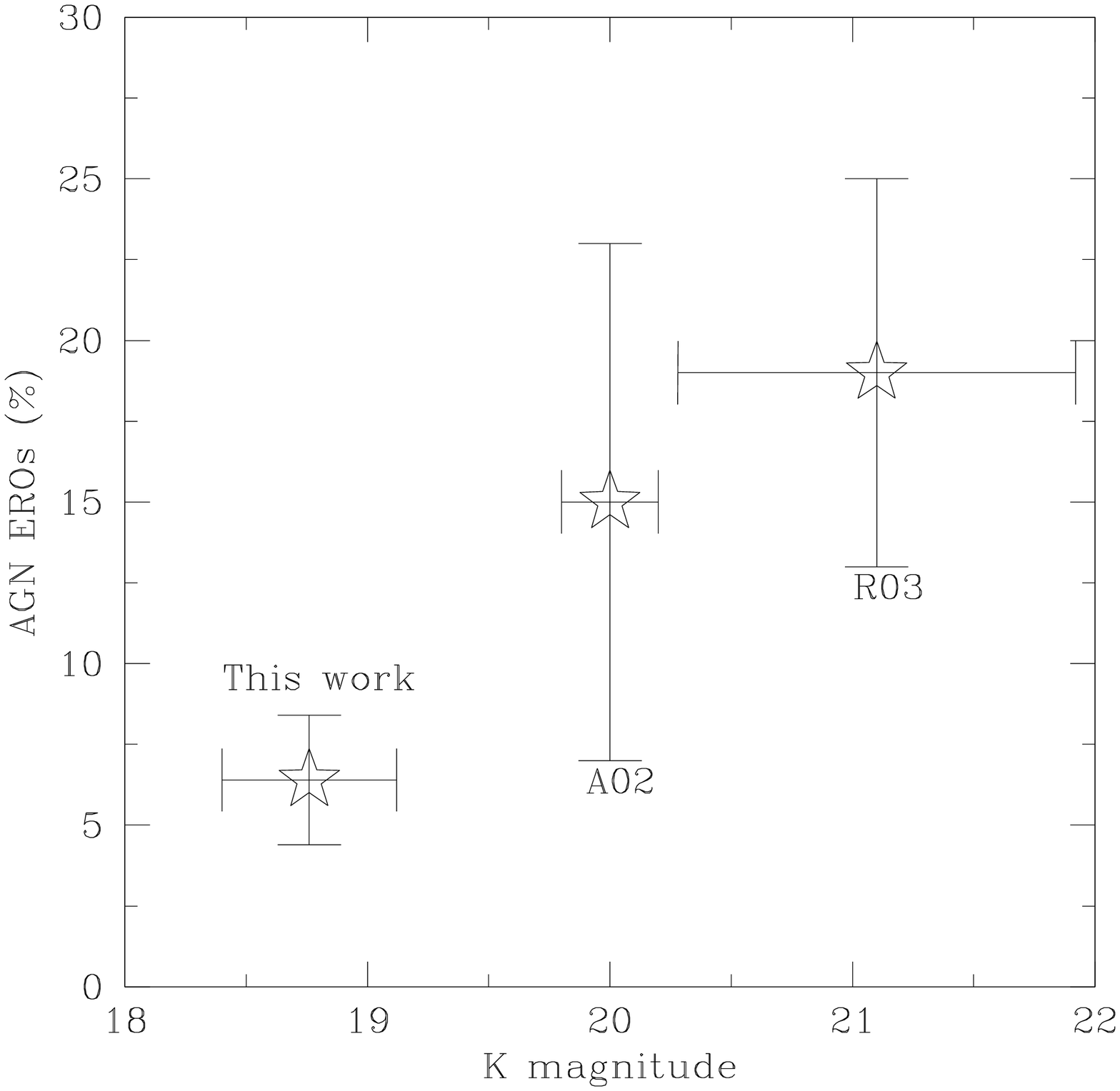}  
\caption{ 
{\it Left panel:} The K magnitude as a function of redshift
for EROs in the literature sample for which spectroscopic (filled
circles) or photometric (empty circles) redshifts are available 
and for near--infrared selected EROS in the K20 survey (crosses:
spectroscopic z; plus signs: photometric z).  
{\it Right panel}: Fraction of AGN ERO as a function of the K--band
magnitudes, as computed from three different samples: this work, the
Alexander et al. (2002) work (A02) and the Roche et al. (2003) sample.
}
\label{kfrac}
\end{figure}

It has been already pointed out, on the basis of a handful of isolated
cases, that the optical and near infrared 
properties of EROs hosting an AGN are indistinguishable
from the overall EROs population both from a spectroscopic and
photometric point of view (Brusa et al. 2002; Cimatti et al. 2003;
Mignoli et al. 2004). 
This seems to apply also to the average redshifts and absolute
luminosities of AGN EROs when compared to the general EROs population.
Figure~\ref{kfrac} (left panel) shows
the K-z plane for the 62 AGN EROs with redshift information (26
spectroscopic and 36 photometric) discussed in the present work 
compared with all the EROs detected in a K--selected survey,
the {\tt K20} survey (Cimatti et al. 2003), for which 
spectroscopic or reliable photometric redshifts are available.
Although this figure shows objects from samples with different $Ks$
limits, it allows us to conclude that X--ray detected EROs appear to be
largely indistinguishable from the general non--AGN EROs population,
both classes being brighter than typical local (z=0) $L^*_K$ galaxies
(Cole et~al.~\cite{cole}; continuous line) and, on average, similar to
evolved $L^*_K$~galaxies at $z=1-1.5$ (Pozzetti et~al. \cite{K20_5};
dashed line), and spanning on average a similar range in the redshift
distribution (z=0.7$\div$2.5). \\
\pn
On the basis of the considerations above, we can conclude that 
the X--ray emission can be considered the only reliable tracer of
AGN activity in the EROs population, for which an accurate analysis 
at longer wavelengths is generally hampered by the faintness of the
optical counterparts.  
Therefore, assuming that the observed distribution of the X/$K$ ratios
(X/K=0.1-10, see Fig.~\ref{rxeros}) is representative of the AGN
EROs population, it is possible to derive an estimate of the
fraction of AGN EROs among $K$--selected samples, as a 
function of the K--band magnitude only and independent from 
the X--ray limiting fluxes. 
Indeed, given that the ratio between the 2--10 keV and $Ks$ limiting
fluxes in our \xmm\ sample is X/K$\sim 1$,  
only about half of the shaded area in the right panel of
Fig.~\ref{rxeros} is sampled at K$\sim 19$, and 
a bias against sources brighter than $Ks\sim 19$ and 
with F$_{2-10 keV}\ls 4\times 10^{-15}$ \cgs\  is
present.  
We have therefore ``corrected'' the estimated fraction of AGN EROs 
reported in Sect.~\ref{frac_eros} taking into account the 
observed distribution of the X--ray to near--infrared ratios:
the fraction of AGN EROs at $K\sim19$ rises up to 7$\pm2$\%. 
In the same way, it is possible to statistically ``correct'' the
observed values in the  A02 and R03 samples already reported in
Sect.~\ref{frac_eros}, in order to derive the AGN fraction at 
K=20.1 and K=21.5, respectively. 
At the limiting fluxes of the A02 sample, the X/K ratio is $\sim
0.2$, while at the R03 sample limiting fluxes, 
the observed X/K is $\sim 1.5$. 
Combining the depth of the two different samples with the 
observed X--ray to optical flux ratios distribution, the estimates of
the fraction of AGN among EROs in the A02 and R03 samples rise up
to (15$\pm8$)\% and (19$\pm$6\%), respectively.  \\
In the right panel of Fig.~\ref{kfrac} these fractions
for the three samples are shown at representative
K--band  magnitudes. 
Even if the statistical error bars are large, 
Figure~\ref{kfrac} suggests that the fraction of AGN EROs among the 
K--selected EROs population is an increasing function of the  
K--band magnitude. 
The results from hard X--ray surveys indicate a space density 
of low--luminosity (10$^{42}-10^{44}$ erg s$^{-1}$) AGN 
almost two order of magnitudes higher than that of high luminosity sources
(Fiore et al. 2003; Ueda et al. 2003).
Thus, it is not surprising that the fraction of AGN EROs increases
going toward faint fluxes (i.e., lower luminosities).
Finally, it is worth remarking that the fraction of ``active'' objects in
K--selected EROs samples can be used to constrain models which link
the formation and evolution of galaxies and AGN (e.g. Granato et
al. 2004).  
\section{X--ray Properties of AGN EROs}
\label{erosx}
In order to check whether X-ray absorption is common among hard X--ray
detected EROs, we have quantitatively measured the intrinsic X--ray
column densities for the 62 EROs with a spectroscopic or
photometric redshift available in the literature sample
described in Sect.~\ref{comparison}. \\
Column densities for the sources detected in the CDF--N and CDF--S have
been obtained by fitting the observed counts with a single power law
model plus absorption at the source redshift.
When the quality of the X-ray spectra in terms of S/N ratio was not
sufficient to use the standard $\chi^{2}$ statistic (a limit of 
150 counts over the 0.5--8 keV band has been assumed), the C-Statistic
was used (Cash 1979).  
In this case the power--law spectral index has been fixed at $\Gamma$=1.9. 
For the sources from the Lockman Hole, the \hel\ and the
``additional'' sample,  the best--fit values quoted by the authors have
been adopted. 
In all the cases, 2--10 keV luminosities were estimated from
the observed  X--ray fluxes and corrected for absorption.\\
The results are reported in Fig.~\ref{nhlum}.
Almost all of the individually detected EROs are consistent with
intrinsic column densities in excess of 10$^{22}$ cm$^{-2}$, and 
they actually {\it are} heavily obscured AGN. 
This study statistically confirms previous evidences, some of which
based on HR analysis (Alexander et al. 2002) and on
spectral analysis (e.g. Vignali et al. 2003; Gandhi et al. 2004; 
Willott et al. 2003; Stevens et al. 2003; Severgnini et al. in prep), 
and unambiguously indicates
that large columns of cold gas (even $> 10^{23}$ cm$^{-2}$) are the rule rather
than the exception in EROs individually detected in the X--rays. 

\subsection{EROs and QSO2: a selection criterion}
\begin{figure}[!t]
\center
\includegraphics[width=0.8\textwidth]{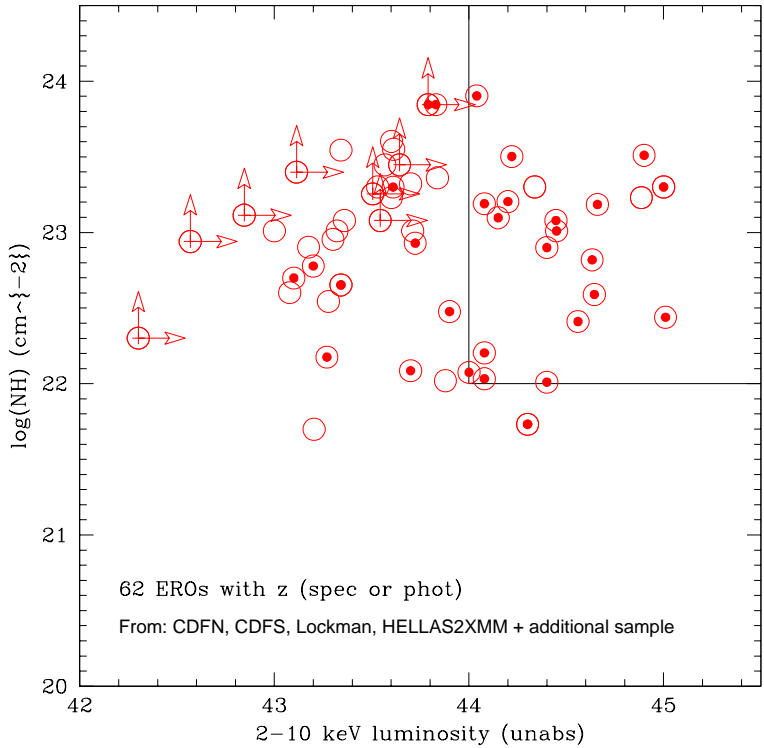}  
\caption{Logarithm of the absorbing column density  (N$_H$) 
versus the logarithm of the unabsorbed X--ray luminosity in the 
2-10 keV band for all the X--ray detected EROs with spectroscopic or 
photometric redshifts from the literature sample. 
Filled symbols are those with X/O $>10$
(see text). In the upper right corner the ``QSO2 locus'' is
highlighted.}
\label{nhlum}
\end{figure}

\pn
Given the high redshift (z$\gsimeq 1$) and the X--ray flux of these objects, 
it follows that the majority of X--ray
detected EROs have high X--ray luminosities (L$_X>10^{43}$ erg
s$^{-1}$, see Fig.~\ref{nhlum}). Moreover, according to our analysis,
a large fraction of the objects analysed in this work
for which redshift information is available have X--ray 
luminosities even larger than $10^{44}$ erg s$^{-1}$, and
therefore well within the quasar regime. 
The large intrinsic column densities further imply that AGN EROs, 
selected at the brightest X--ray fluxes, have properties 
similar to those of quasars 2 (QSO2), the high--luminosity, 
high redshift type II AGNs predicted by X--ray background synthesis
models and necessary to reproduce the 2--10 keV source counts 
at relatively bright fluxes (e.g. Comastri et al. 2001; Gilli, Salvati
\& Hasinger 2001). \\
On the basis of unified schemes, type II quasars 
are expected to be luminous, narrow--line, high--redshift objects
with substantial (N$_H>10^{22}$ cm$^{-2}$) X--ray column
densities. However, at the faint 
fluxes/magnitudes of deep surveys the optical identifications of these
objects is very difficult, unless one of the strong emission lines is
present in the optical spectrum. 
Our analysis suggests that an efficient method to pick up
this elusive population is the combination of 
medium--deep X--ray observations and K--band imaging: 
among hard X--ray sources, one must select those counterparts 
with an R$-$K$>5$ color --- that is an indication of high redshift and
obscuration --- and with an X/O ratio $>10$ --- 
that is an indication of high column densities and high luminosity
(see also Severgnini et al. 2004; Gandhi et al. 2004). 
This is shown in Fig.~\ref{nhlum}, where EROs with
X/O$>10$ are reported as filled symbols 
and populate the upper right region of the diagram
(the ``QSO2 locus'').
The present work therefore confirms that a selection on the basis of
X/O$>10$ is a powerful tool to detect high--luminosity, highly
obscured sources as already pointed out by Fiore et al. (2003),  
and it is even stronger when coupled with a previous selection on the
basis of extremely red colors. 
\pn
It is important to stress that 
EROs with high X/O can be QSO2, but it is not true that {\it all} the QSO2
are EROs.  As an example, the prototype of high--redshift QSO2,
CDFS\_202 in Norman et al. (2002), has R$-$K$\sim$2.5.
In this case the rather blue observed colour is mainly due to the
presence of a strong emission line in the R filter.
When the line flux is
subtracted, the R$-$K colour of CDFS\_202 is R$-$K$\sim$4, which makes
this object appreciably red. \\
\pn
The close link between X--ray bright EROs and type 2 quasars
allows us to elaborate on the contribution of EROs to 
the population of high luminosity, highly obscured quasars. 
We have considered only the sources with redshift
information detected in the CDFN, CDFS, and in the Lockman Hole. In
addition, using the method
extensively discussed by Fiore et al. (2003) 
based on the relation between the X/O and the X--ray luminosity,
we have derived the redshifts and luminosities also for the 9 hard
X--ray detected EROs in our \xmm\ observation. 
All but two have unabsorbed X--ray luminosities larger than 10$^{44}$ erg
s$^{-1}$ and lie in the redshift range z=1-3. \\
To estimate the contribution of EROs to the QSO2 population we
have chosen a 2--10 keV limiting flux of $\sim 10^{-14}$
\cgs, in order to ensure a flat and uniform
sky coverage down to fluxes where about half of the XRB flux is
resolved.  
The most recent published determinations of the space density of type 2
quasars at fluxes brighter than this limit are in the range 40--50
deg$^{2}$ (Perola et al. 2004; Padovani et al. 2004) while the
prediction from the Ueda et al. (2003) model  is $\sim 75$ deg$^{-2}$
(private communication). 
From our analysis, a total of 6 EROs over an area of about 0.4
deg$^{2}$ have been detected at fluxes $\gs 10^{-14}$ \cgs\ and 
classified as QSO2, i.e. they have $N_H > 10^{22}$ cm$^{-2}$ and 
unabsorbed L$_{2-10 \rm keV}>$
10$^{44}$ erg s$^{-1}$.
Therefore, the surface density of luminous, obscured EROs 
is about 15 deg$^{-2}$, and it has to be regarded as a robust lower
limit given the lack of redshift information for some of the EROs in 
the present sample.
This work therefore indicates that AGN EROs represent {\it at least}
20\% of the type 2 quasars population, if compared to the Ueda et
al. (2003) predictions, and it can be as high as $\sim$40\% when compared
with the recently published estimates (Padovani et al. 2004; Perola et
al. 2004). \\

\subsection{X/K correlation and the accretion parameters of AGN EROs}

\begin{figure}[!t]
\center
\includegraphics[width=0.98\textwidth]{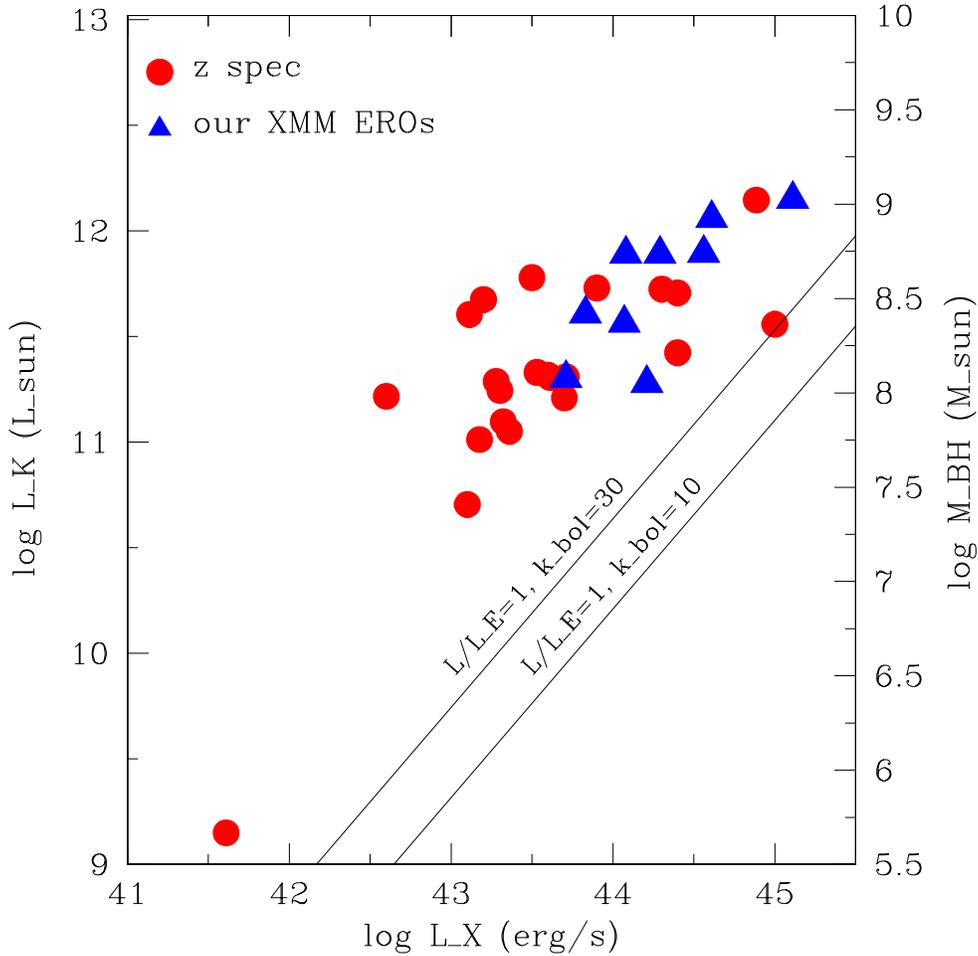}  
\caption{
K--band luminosities versus the 2--10 keV X--ray
luminosities (filled triangles: our XMM EROs; filled circles: 
EROs with spectroscopic redshifts). 
The M$_{BH}$ resulting from the K--band luminosity on the basis of the 
Marconi \& Hunt (2003) relation are reported in the right axis
of the Figure.
The two continuous lines
represent the expected correlation between the two plotted quantities
for two different assumptions on the bolometric correction
($k_{bol}$=10 and $k_{bol}$=30) and for L/L$_{Edd}$=1.}
\label{masses}
\end{figure}

While there is not a clear trend between the X--ray flux 
and optical magnitude (left panel of Figure~\ref{rxeros}),
a linear correlation 
characterized by a relatively small scatter 
appears to be present between the X--ray and the K--band fluxes
(right panel of Figure~\ref{rxeros}).
This relation is present despite the large redshift range 
(e.g. $z\simeq0-3$) of our sources 
and may reflect a proportionality also 
between the luminosities, $L_X \propto L_K$.
Such a correlation is reminiscent of the ones observed 
locally between the BH mass and the global 
galaxy properties (Magorrian et al.
1998; Gebhardt et al. 2000; Ferrarese \& Merritt 2000; Marconi \& Hunt
2003) and may be related to them. \\
In order to test such a possibility with a conservative approach 
we have considered those EROs with a secure spectroscopic identification in 
the comparison sample and the 9 EROs in our XMM--{\it Newton} 
observation, using the Fiore et al. (2003) relation to 
estimate their redshifts.
Although EROs and bluer AGN have similar
X/K ratio, a result somehow surprising given that both
the active nucleus and the host galaxies are contributing to the
$K$-band light presumably in different ratios, we limit our analysis to EROs
since there are several indications that the near--infrared emission of 
these X--ray selected {\it obscured} AGN is dominated by their host 
galaxy starlight (see e.g. Mainieri et al. 2002; Mignoli et al. 2004).
The rest--frame K--band luminosities have been computed
using an evolving {\it elliptical} galaxy template 
to properly account for the K--corrections\footnote{
Note that in the K--band the K--corrections are relatively insensitive to
galaxy type and fairly small up to z$\ls2$.} (Bruzual \& Charlot 2003).
The results are shown in Fig.~\ref{masses}. \\
The correlation between the near--infrared and
X--ray luminosities observed for these high--redshift EROs 
(i.e. $L_X \propto L_K$) closely 
resembles the one recently published by Marconi \& Hunt (2003) 
between the BH mass and the K band luminosity 
for a sample of local galaxies. 
 Assuming that the X--ray luminosity of our AGN EROs is proportional 
to the BH mass (i.e. the Eddington ratio L/L$_{\rm Edd}$ and the bolometric
correction $k_{\rm bol}$\footnote{The absorption corrected X--ray 
luminosity can be translated into a bolometric luminosity assuming 
a bolometric correction factor (L$_{bol}=k_{bol}\times L_X$).}
are not a strong function of the BH mass)
the observed correlation plotted in Fig.~\ref{masses} 
implies L$_{\rm K} \propto {\rm M}_{\rm BH}$. 
It is thus formally possible to
tentatively constrain the BH masses and the accretion parameters that
would follow if also the normalization of such relation does not 
significantly evolve with redshift, if not for the expected change of
stellar mass to light ratio of the host galaxies due to the evolution
of the stellar populations (our AGN EROs sample being at $z=1$--2).
The resulting M$_{BH}$ are reported in the right axis of 
Fig.~\ref{masses}.
The two continuous lines in Fig.~\ref{masses} represent
the relation between the BH mass and X--ray luminosity
computed for Eddington limited accretion ($L_{bol}/L_{edd}=1$) 
and for two
different values ($k_{bol}$=30 and $k_{bol}$=10) of the bolometric
correction. The first value is from the Quasar SED compilation of 
Elvis et al. (1994) and can be considered a reliable correction 
for bright unobscured quasars.
The second appears to be appropriate for lower luminosities (L$_{\rm X}\sim 
10^{43}$-10$^{44}$ erg s$^{-1}$) Seyfert like galaxies (Fabian 2004) 
and a few heavily obscured, luminous sources (Comastri 2004).  
The value observed for our objects can be reasonably well explained by 
a spread in the Eddington ratios in the range 
$L_{\rm bol}/L_{\rm Edd}=2\times10^{-3}\div1$, with a median value of 
$L_{\rm bol}/L_{\rm Edd}=0.03-0.1$ for k$_{\rm bol}=10$ and 30, respectively.
Both the BH masses and the Eddington ratios derived above  
are consistent with a scenario in which X--ray detected 
EROs are obscured quasars emitting in a radiatively efficient way, 
in agreement with the results 
of Merloni (2004) and McLure \& Dunlop (2004). 
These findings are broadly consistent with those obtained, with completely
different methods by Woo and Urry (2002) for a large sample of broad
line AGN (see their Figure 8). \\
 Although the discussion above has been conducted only in a
qualitative way and neglecting the uncertainties and scatter
associated to the observed relations, it appears fully reasonable 
that the close correlation that we have detected between the X--ray flux and 
the K--band magnitude of AGN EROs is the high redshift analogous 
of the correlations observed locally,  
implying a close connection between the black hole
and AGN host galaxies also at earlier epochs ($z=1-2$).

\section{Summary}
In this paper we have presented the results from an 80 ks \xmm\
observation of the largest sample of near--infrared selected
EROs available to date down to a K--band magnitude limit of K=19.2
(Daddi et al. 2000).  
The moderate--deep exposure and the high energy throughput of
\xmm, coupled with its large field of view, allowed us to detect,
for the first time on a statistically significant sample, 
AGN--powered EROs at relatively bright X--ray
fluxes.
At the limiting fluxes probed by our survey (F$_{2-10
keV}\gs4\times10^{-15}$ and $Ks\ls 19.2$) the fraction of AGN EROs
within near--infrared selected ERO samples is $\sim$3.5\%. 
Conversely, a significant fraction of the optical counterparts of hard X--ray
selected sources are EROs and the fraction of extremely red objects among 
the X--ray population is much higher ($\sim 20$\%). \\
The average hardness ratio of the hard X--ray detected EROs in the
\xmm\ observation suggest substantial column densities at the source
redshift. \\
In order to place our results in a broader context, 
we have also considered additional samples of X--ray detected EROs
available in the literature from published deep
and medium deep hard X--ray surveys. 
A total of 128 X--ray detected EROs have been considered, and  
for the first time the average X--ray, optical and near--infrared
properties of AGN--powered EROs have been derived on a statistically
significant sample.  
The most important results of our analysis are summarized in the following: 
\begin{itemize}
\item[$\bullet$]  
The average X--ray to optical flux ratio of AGN EROs
is about one order of magnitude larger than that
observed for BL AGN. 
On the contrary, when the X--ray to near--infrared
properties are considered, all the EROs in the comparison sample
occupy a locus which is indistinguishable from that occupied by 
unobscured QSO.This results further corroborates the hypotheses 
that AGN EROs are obscured quasars. 
\item[$\bullet$]   
High obscuration in X--ray detected EROs is also revealed in
the X--ray band, in agreement with the results from the optical band:
the majority of the sources with known redshifts of the comparison
sample have N$_H>10^{22}$ cm$^{-2}$, and about half even in excess
than 10$^{23}$ cm$^{-2}$. 
The observed X--ray fluxes and spectral shapes imply unabsorbed, rest
frame X--ray luminosities in the range 10$^{42}$--10$^{45}$ erg s$^{-1}$.  
At fluxes larger than 10$^{-14}$ \cgs, we estimate that AGN EROs
contribute at least 20\% (and possibly up to 40\%) to the QSO2
population. 
A selection criterion based on the X/O and the R$-$K 
colour of hard X--ray selected sources has been proposed, to efficiently 
pick--up this elusive population of highly obscured quasars that are
one of the key ingredient of XRB synthesis models.
\item[$\bullet$] 
Assuming that the observed X/K distribution of EROs is representative 
of the AGN EROs population, and combining our results with 
those of A02 and R03, 
we found an evidence of an increase of the fraction of AGN EROs in
K--selected EROs samples as a function of the K--band magnitude,
ranging from $\sim 6$\% at $K$=19 to $\sim 20$\% at $K=22$.
\item[$\bullet$]  
The close relation uncovered between the K- and X-ray band fluxes of
AGN EROs suggests that the connection between 
the properties of the host galaxies and central black holes 
 observed locally holds also at higher redshifts ($z \simeq 1-2$). 
This allows an estimate of the BH masses and Eddington ratios  
for the sources with known redshift through
reasonable assumptions. The results indicate that the majority of
AGN powered EROs have BH masses larger than 5$\times 10^{7}$ \msun, 
and are accreting with a median value of L$_{\rm bol}$/L$_{\rm Edd} 
\sim 0.03-0.1$.
\end{itemize}
\pn
All the findings discussed above support the idea that 
hard X--ray surveys coupled with near--infrared observations 
provide an efficient method in detecting QSO2. Furthermore,  
X--ray detected EROs can be used as lighthouses to 
investigate the accretion paradigm at high redshifts to address the 
issue of elliptical galaxy formation and the expected co-evolution
with the accreting black-holes.
In particular, systematic studies of the relationship
between EROs and QSO2 are needed 
to {\it quantitatively} investigate the
link between the formation of massive elliptical galaxies and the
onset of AGN activity. \\ 
X--ray observations of large samples
of $K$--selected EROs would be crucial to compute the fraction of 
X--ray active EROs on the widest area possible (to avoid cosmic
variance). Conversely, deep optical and near--infrared follow--up of complete
samples of hard X--ray selected sources with extreme X/O
will definitively assess the fraction of reddened
sources among the XRB constituents. 
The full exploitation of the COSMOS multiwavelength database
will be in the near future the best 
strategy to investigate both these issues. The large
area covered ($\sim 2$ deg$^{2}$) will allow a detailed study of the
clustering properties of these objects and would shed new light on the
link between nuclear activity and galaxy evolution.\\

\begin{acknowledgements}
\par\noindent
We gratefully acknowledge the K20 team for providing their data, and
Y. Ueda for providing his model predictions.
MB and AC kindly acknowledge support 
by INAOE, Mexico, during the 2003 Guillermo-Haro Workshop where part 
of this work was performed.
MB acknowledges partial support from the Deutscher Akademischer
Austausch Dienst (DAAD, German Academic Exchange Service) under the project:
``New Frontiers in Science''.
The XMM-{\it Newton} Helpdesk and the entire \xmm\ team, in particular
B. Altieri, M. Guainazzi and G. Vacanti, are kindly acknowledged for
their help with astrometry problems. 
The authors acknowledge partial support by ASI I/R/057/02 and 
MIUR COFIN--03--02--23 contracts, and INAF 270/2003 grant.
\end{acknowledgements}

\begin{table*}
       \caption{Properties of the X--ray selected sources}
       \label{tab+xray}  
       \vspace{0.2cm}
\begin{minipage}{0.99\textwidth}
\small
\begin{tabular}{ccccccc}
\hline\hline\noalign{\smallskip}                                              
ID     &
RA       &    
DEC      &
2--10 keV flux &
hard counts  &
0.5--2 keV flux & 
soft counts \\
 & (J2000) & (J2000) & (erg cm$^{-2}$ s$^{-1}$) & & (erg cm$^{-2}$ s$^{-1}$) & \\ 
\noalign{\smallskip}\hline\noalign{\smallskip} 	
 65 & 222.40211 &   9.14214 & 1.07e-13 & 823.4 $\pm$  38.1 & 7.07e-14 & 2685.5 $\pm$ 	 65.2 	 \\
246 & 222.52921 &   8.99011 & 5.54e-14 & 365.0 $\pm$  27.4 & 4.24e-14 & 1401.9 $\pm$ 	 48.4 	 \\
348 & 222.21439 &   8.89961 & 5.49e-14 & 350.6 $\pm$  27.0 & ...      &  $<47.7$ \\
240 & 222.43640 &   8.99042 & 5.38e-14 & 524.5 $\pm$  31.3 & 4.03e-14 & 1873.8 $\pm$ 	 55.1 	 \\
300 & 222.29083 &   8.87070 & 4.58e-14 & 361.0 $\pm$  27.2 & 1.31e-14 & 508.1 $\pm$ 		 30.9 	\\
321 & 222.42348 &   8.92257 & 3.93e-14 & 398.8 $\pm$  27.9 & 7.01e-15 & 349.3 $\pm$ 		 26.7 	\\
226 & 222.35175 &   8.99408 & 3.13e-14 & 424.2 $\pm$  28.6 & 1.95e-14 & 1273.3 $\pm$ 	 46.4 	 \\
 72 & 222.24762 &   9.12793 & 2.93e-14 & 205.8 $\pm$  22.3 & 7.03e-15 & 243.7 $\pm$ 		 23.5 	\\
148 & 222.40685 &   9.05575 & 2.89e-14 & 332.3 $\pm$  26.0 & 1.52e-14 & 848.8 $\pm$ 		 38.3 	\\
296 & 222.36632 &   8.94443 & 2.70e-14 & 328.0 $\pm$  25.7 & 1.01e-14 & 595.8 $\pm$ 		 33.0 	\\
247 & 222.49405 &   8.98667 & 2.62e-14 & 214.5 $\pm$  22.5 & 1.44e-14 & 585.8 $\pm$ 		 32.7 	\\
195 & 222.23718 &   9.02275 & 2.52e-14 & 228.2 $\pm$  22.9 & 5.13e-15 & 225.8 $\pm$ 		 23.0 	\\
338 & 222.44652 &   8.90643 & 2.50e-14 & 175.3 $\pm$  20.7 & 1.92e-14 & 643.5 $\pm$ 		 34.3 	\\
258 & 222.16916 &   8.97851 & 2.02e-14 & 119.7 $\pm$  19.4 & 1.75e-14 & 515.0 $\pm$ 		 31.4 	\\
 67 & 222.40207 &   9.13641 & 2.02e-14 & 146.0 $\pm$  19.7 & 1.60e-14 & 571.8 $\pm$ 		 30.4 	\\
138 & 222.26871 &   9.06119 & 1.67e-14 & 157.3 $\pm$  19.8 & 9.36e-15 & 426.3 $\pm$ 		 28.9 	\\
369 & 222.41972 &   8.87781 & 1.61e-14 & 115.2 $\pm$  18.4 & 6.06e-15 & 218.8 $\pm$ 		 22.3 	\\
370 & 222.46298 &   8.87898 & 1.55e-14 & 110.3 $\pm$  18.7 & 9.65e-15 & 343.7 $\pm$ 		 26.4 	\\
254 & 222.45338 &   8.98117 & 1.54e-14 & 127.7 $\pm$  18.5 & 1.14e-14 & 454.5 $\pm$ 		 29.3 	\\
217 & 222.45128 &   8.99983 & 1.48e-14 & 156.2 $\pm$  19.6 & 9.91e-16 &  51.1 $\pm$ 		 15.2 	\\
170 & 222.27676 &   9.04376 & 1.40e-14 & 144.9 $\pm$  19.5 & 6.31e-15 & 314.7 $\pm$ 		 25.6 	\\
269 & 222.24794 &   8.96716 & 1.37e-14 & 128.3 $\pm$  19.2 & 8.73e-15 & 398.3 $\pm$ 		 28.0 	\\
337 & 222.35938 &   8.90312 & 1.35e-14 & 102.2 $\pm$  17.3 & 5.91e-15 & 223.4 $\pm$ 		 22.8 	\\
177 & 222.48996 &   9.03878 & 1.33e-14 &  95.3 $\pm$  16.9 & 1.13e-14 & 393.5 $\pm$ 		 27.4 	\\
169 & 222.35056 &   8.83613 & 1.21e-14 &  74.8 $\pm$  16.7 & 4.76e-15 & 143.8 $\pm$ 		 19.5 	\\
350 & 222.28494 &   8.89601 & 1.17e-14 &  97.3 $\pm$  17.4 & 7.32e-15 & 296.4 $\pm$ 		 25.2 	\\
361 & 222.40059 &   8.89156 & 1.15e-14 &  96.0 $\pm$  17.3 & ... & $<42.0$                    	\\
326 & 222.31322 &   8.87259 & 1.11e-14 &  90.7 $\pm$  17.7 & 3.80e-15 & 153.3 $\pm$ 		 20.5 	\\
367 & 222.39218 &   8.88237 & 1.02e-14 &  91.5 $\pm$  17.6 & 4.84e-15 & 213.2 $\pm$ 		 22.5 	\\
 66 & 222.41061 &   9.13401 & 1.02e-14 &  81.4 $\pm$  16.7 & 6.70e-15 & 262.6 $\pm$ 		 24.1 	\\
150 & 222.29356 &   9.05019 & 1.00e-14 & 110.7 $\pm$  18.3 & ...      & $<43.8$ \\
219 & 222.49629 &   8.99483 & 9.95e-15 &  80.7 $\pm$  16.9 & 4.52e-15 & 181.8 $\pm$ 		 20.9 	\\
209 & 222.39635 &   9.00872 & 9.48e-15 & 121.3 $\pm$  18.1 & 6.61e-15 & 408.4 $\pm$ 		 27.8 	\\
293 & 222.35324 &   8.94158 & 9.13e-15 & 109.7 $\pm$  17.7 & ... &  $<48.3$ \\                      
181 & 222.45479 &   9.03205 & 9.01e-15 &  90.8 $\pm$  17.1 & 2.19e-15 & 108.0 $\pm$ 		 18.3 	\\
239 & 222.30937 &   8.85337 & 8.90e-15 &  63.4 $\pm$  16.7 & 2.63e-15 &  93.1 $\pm$ 		 17.8 	\\
362 & 222.43134 &   8.88969 & 8.86e-15 &  69.1 $\pm$  15.9 & 1.89e-15 &  73.9 $\pm$ 		 16.7 	\\
 23 & 222.37926 &   9.17607 & 8.59e-15 &  54.5 $\pm$  16.4 & 1.72e-15 &  53.9 $\pm$ 		 16.1 	\\
328 & 222.49466 &   8.91315 & 8.15e-15 &  57.8 $\pm$  16.3 & ... & $<42.9$ \\
255 & 222.35718 &   8.97293 & 8.03e-15 & 105.8 $\pm$  17.6 & 6.89e-15 & 438.1 $\pm$ 		 29.6 	\\
310 & 222.38539 &   8.87228 & 8.02e-15 &  68.0 $\pm$  16.4 & 4.95e-15 & 207.2 $\pm$ 		 22.1 	\\
288 & 222.33345 &   8.94491 & 8.00e-15 &  95.1 $\pm$  17.3 & 2.97e-15 & 171.2 $\pm$ 		 20.8 	\\
175 & 222.32428 &   9.03741 & 7.91e-15 &  98.6 $\pm$  17.5 & 1.30e-15 &  77.9 $\pm$ 		 16.6 	\\
\noalign{\smallskip}\hline
\end{tabular}
\end{minipage}
\end{table*}                                                                   
\newpage
\begin{table*}
\begin{minipage}{0.99\textwidth}

\small
\begin{tabular}{ccccccc}
\hline\hline\noalign{\smallskip}                                              
ID     &
RA       &    
DEC      &
2--10 keV flux &
hard counts  &
0.5--2 keV flux &
soft counts  \\
 & (J2000) & (J2000) & (erg cm$^{-2}$ s$^{-1}$) & & (erg cm$^{-2}$
s$^{-1}$) & \\ 
\noalign{\smallskip}\hline\noalign{\smallskip} 	
360 & 222.33759 &   8.88898 & 7.85e-15 &  73.7 $\pm$  16.6 & ... & $<39.9$ \\
357 & 222.42833 &   8.88944 & 7.54e-15 &  56.0 $\pm$  15.2 & 2.30e-15 &  86.2 $\pm$ 		 17.1 	\\
152 & 222.36366 &   9.05004 & 7.49e-15 &  93.9 $\pm$  17.1 & ... & $<44.1$ \\
330 & 222.45470 &   8.91179 & 7.43e-15 &  58.1 $\pm$  15.7 & 2.91e-15 & 111.0 $\pm$ 		 18.2 	\\
168 & 222.26262 &   9.04336 & 7.20e-15 &  61.4 $\pm$  15.1 & 3.36e-15 & 136.8 $\pm$ 		 18.7 	\\
186 & 222.39972 &   9.03090 & 7.19e-15 &  89.4 $\pm$  16.8 & 4.66e-15 & 280.5 $\pm$ 		 25.5 	\\
244 & 222.26265 &   8.98473 & 7.10e-15 &  63.9 $\pm$  16.1 & 2.13e-15 &  94.5 $\pm$ 		 17.6 	\\
178 & 222.27762 &   9.03575 & 6.99e-15 &  61.0 $\pm$  15.2 & 6.03e-15 & 257.4 $\pm$ 		 23.4 	\\
249 & 222.29012 &   8.97873 & 6.57e-15 &  75.3 $\pm$  16.1 & 3.80e-15 & 210.7 $\pm$ 		 22.3 	\\
146 & 222.41490 &   9.05666 & 6.48e-15 &  67.2 $\pm$  14.9 & 1.32e-15 &  65.9 $\pm$ 		 16.1 	\\
210 & 222.39804 &   9.01260 & 6.01e-15      &  76.0$\pm$16.8          & 3.51e-15 & 216$\pm$22.6 \\                          
189 & 222.26312 &   9.02621 & 5.82e-15 &  56.8 $\pm$  15.3 & ... & $<41.7$ \\
250 & 222.43520 &   8.97954 & 5.77e-15 &  65.0 $\pm$  15.9 & ... & $<45.0$ \\
154 & 222.39172 &   9.05019 & 5.36e-15 &  64.7 $\pm$  16.0 & ... & $<43.8$ \\
263 & 222.36026 &   8.97151 & 4.97e-15 &  65.1 $\pm$  15.9 & 6.59e-15 & 417.1 $\pm$ 		 29.1 	\\
176 & 222.42131 &   9.03696 & 4.94e-15 &  48.4 $\pm$  13.4 & 4.02e-15 & 186.4 $\pm$ 		 21.1 	\\
251 & 222.29750 &   8.97501 & 4.61e-15 &  53.8 $\pm$  14.7 & ... & $<45.0$ \\
287 & 222.29202 &   8.94793 & ...      &  $<43.2$          & 1.73e-14 & 917.7 $\pm$ 		 40.0 	\\
130 & 222.53807 &   9.07928 & ...      &  $<45.9$          & 9.39e-15 & 270.6 $\pm$ 		 25.0 	\\
315 & 222.23370 &   8.92656 & ...      &  $<43.5$          & 7.25e-15 & 277.1 $\pm$ 		 24.3 	\\
207 & 222.48463 &   9.01196 & ...      &  $<39.6$          & 4.66e-15 & 173.4 $\pm$ 		 20.4 	\\
179 & 222.17657 &   9.03417 & ...      &  $<42.9$          & 3.86e-15 & 118.9 $\pm$ 		 18.2 	\\
242 & 222.35513 &   8.98782 & ...      &  $<42.3$          & 3.73e-15 & 242.6 $\pm$ 		 24.0 	\\
344 & 222.23007 &   8.90152 & ...      &  $<46.5$          & 3.49e-15 & 120.8 $\pm$ 		 19.1 	\\
236 & 222.32793 &   8.85341 & ...      &  $<43.8$          & 3.10e-15 & 116.4 $\pm$ 		 18.3 	\\
118 & 222.18839 &   9.08308 & ...      &  $<30.0$          & 3.05e-15 &  55.2 $\pm$ 		 14.5 	\\
 68 & 222.31238 &   9.12831 & ...      &  $<39.6$          & 3.01e-15 & 123.7 $\pm$ 		 18.9 	\\
103 & 222.27232 &   9.09741 & ...      &  $<42.6$          & 2.95e-15 & 112.3 $\pm$ 		 18.1 	\\
 94 & 222.43176 &   9.10102 & ...      &  $<39.9$          & 2.81e-15 & 106.3 $\pm$ 		 17.6 	\\
363 & 222.46669 &   8.88736 & ...      &  $<45.6$          & 2.71e-15 &  98.7 $\pm$ 		 17.9 	\\
157 & 222.23582 &   9.04881 & ...      &  $<38.7$          & 2.45e-15 &  93.6 $\pm$ 		 18.1 	\\
289 & 222.44258 &   8.85969 & ...      &  $<39.9$          & 2.39e-15 &  83.0 $\pm$ 		 16.9 	\\
273 & 222.18602 &   8.96340 & ...      &  $<43.2$          & 2.35e-15 &  79.0 $\pm$ 		 17.2 	\\
301 & 222.32642 &   8.93808 & ...      &  $<39.9$          & 2.29e-15 & 127.4 $\pm$ 		 18.3 	\\
125 & 222.37196 &   9.07913 & ...      &  $<41.1$          & 2.27e-15 & 123.3 $\pm$ 		 19.2 	\\
259 & 222.34473 &   8.97245 & ...      &  $<43.2$          & 2.26e-15 & 141.7 $\pm$ 		 20.5 	\\
198 & 222.49577 &   9.01764 & ...      &  $<39.0$          & 2.07e-15 &  59.0 $\pm$ 		 15.4 	\\
 77 & 222.34190 &   9.11713 & ...      &  $<40.5$          & 2.05e-15 &  92.5 $\pm$ 		 17.6 	\\
139 & 222.39677 &   9.06015 & ...      &  $<41.4$          & 2.03e-15 & 114.3 $\pm$ 		 18.8 	\\
107 & 222.23544 &   9.09159 & ...      &  $<46.2$          & 2.02e-15 &  77.3 $\pm$ 		 16.9 	\\
 28 & 222.27783 &   9.16305 & ...      &  $<43.2$          & 2.01e-15 &  65.5 $\pm$ 		 16.5 	\\
298 & 222.30969 &   8.94036 & ...      &  $<44.7$          & 1.96e-15 & 106.5 $\pm$ 		 18.3 	\\
319 & 222.28975 &   8.92027 & ...      &  $<41.7$          & 1.84e-15 &  71.6 $\pm$ 		 16.1 	\\
180 & 222.31946 &   9.03277 & ...      &  $<42.0$          & 1.82e-15 &  92.5 $\pm$ 		 17.5 	\\
\noalign{\smallskip}\hline
\end{tabular}
\end{minipage}
\end{table*}                                                                   
\newpage
\begin{table*}
\begin{minipage}{0.99\textwidth}

\small
\begin{tabular}{ccccccc}
\hline\hline\noalign{\smallskip}                                              
ID     &
RA       &    
DEC      &
2--10 keV flux &
hard counts  &
0.5--2 keV flux &
soft counts  \\
 & (J2000) & (J2000) & (erg cm$^{-2}$ s$^{-1}$) & & (erg cm$^{-2}$
s$^{-1}$) & \\ 
\noalign{\smallskip}\hline\noalign{\smallskip} 	
203 & 222.31351 &   9.01312 & ...      &  $<41.7$          & 1.72e-15 & 103.6 $\pm$ 		 17.3 	\\
 73 & 222.26277 &   9.12213 & ...      &  $<38.4$          & 1.64e-15 &  60.9 $\pm$ 		 15.6 	\\
122 & 222.48824 &   9.08229 & ...      &  $<45.3$          & 1.63e-15 &  58.4 $\pm$ 		 15.8 	\\
158 & 222.25018 &   9.04792 & ...      &  $<37.2$          & 1.45e-15 &  56.6 $\pm$ 		 15.2 	\\
 61 & 222.33221 &   9.13775 & ...      &  $<42.3$          & 1.43e-15 &  57.5 $\pm$ 		 15.7 	\\
282 & 222.22620 &   8.94736 & ...      &  $<43.6$          & 1.41e-15 &  49.5 $\pm$ 		 15.2 	\\
303 & 222.36115 &   8.93673 & ...      &  $<39.6$          & 1.38e-15 &  79.3 $\pm$ 		 17.7 	\\
129 & 222.40413 &   9.07376 & ...      &  $<42.6$          & 1.32e-15 &  70.3 $\pm$ 		 16.3 	\\
241 & 222.21678 &   8.98637 & ...      &  $<45.9$          & 1.27e-15 &  50.5 $\pm$ 		 15.6 	\\
237 & 222.33328 &   8.98678 & ...      &  $<38.4$          & 1.03e-15 &  65.0 $\pm$ 		 16.3 	\\
\noalign{\smallskip}\hline
\end{tabular}
\end{minipage}
\vspace{0.5cm}
Notes: ID = X--ray source identifications; RA, DEC = X--ray
coordinates.
\end{table*}

\begin{table*}
       \caption{Optical and near--infrared photometry of the selected sources}
       \label{tab_xray_opt}
       \vspace{0.2cm}
\begin{minipage}{0.99\textwidth}
\footnotesize
\begin{tabular}{cccccccc}
\hline\hline\noalign{\smallskip}                                              
ID     &
RA(X)       &    
DEC(X)      &
$\Delta$(X-O) &
R      &
K      &     
LR(R)      &
LR(K)    \\
\noalign{\smallskip}\hline\noalign{\smallskip} 	
 65 & 222.40211 &   9.14214 & 1.68 & 19.74 & 17.94 &   39.98 &   15.98 \\
246 & 222.52922 &   8.99011 & 2.22 & 19.55 & 17.15 &   29.14 &   13.19 \\
348 & 222.21437 &   8.89961 & 0.60 & 21.37 & 16.93 &   10.64 &   24.27 \\
240 & 222.43642 &   8.99043 & 1.37 & 18.48 & 15.48 &  102.81 &   17.78 \\
300 & 222.29082 &   8.87071 & 0.71 & 19.55 & 16.72 &   56.62 &   23.78\\
321 & 222.42348 &   8.92257 & 0.83 & 21.46 & 17.37 &   10.13 &   24.93\\
226 & 222.35179 &   8.99408 & 0.17 & 21.21 & 18.39 &   15.81 &   19.31 \\
 72 & 222.24759 &   9.12793 & 2.72 & 21.73 & 17.43 &    3.70 &    9.10\\
148 & 222.40681 &   9.05575 & 0.37 & 24.71 & 18.72 &    0.12 &   15.78 \\
296 & 222.36630 &   8.94443 & 0.06 & 21.12 & 17.34 &   15.87 &   27.63 \\
247 & 222.49402 &   8.98667 & 1.26 & 20.52 & 17.35 &   20.30 &   21.78 \\
195 & 222.23720 &   9.02275 & 1.73 & 22.63 & 17.86 &    2.42 &   15.58 \\
338 & 222.44647 &   8.90643 & 1.39 & 20.62 & 18.86 &   19.28 &   12.05 \\
258 & 222.16920 &   8.97851 & 0.31 & 19.57 & 17.55 &   60.51 &   22.52 \\
369 & 222.41971 &   8.87782 & 0.22 & 22.78 & 18.78 &    3.77 &   15.99 \\
370 & 222.46300 &   8.87898 & 1.21 & 20.96 & 18.87 &   12.75 &   12.93 \\
254 & 222.45340 &   8.98117 & 1.30 & 21.16 & 17.63 &   12.32 &   21.45 \\
217 & 222.45132 &   8.99983 & 1.04 & 23.30 & $>19.2$ &    0.65 &    0.00 \\
170 & 222.27676 &   9.04376 & 1.10 & 21.84 & 17.52 &    5.27 &   23.05 \\
337 & 222.35941 &   8.90313 & 1.10 & 22.79 & 18.38 &    3.17 &   16.17 \\
177 & 222.48997 &   9.03878 & 2.28 & 20.66 & 17.68 &   11.81 &   11.19 \\
169 & 222.35057 &   8.83613 & 1.49 & 23.71 & $>19.2$ &    1.18 &    0.00 \\
350 & 222.28493 &   8.89601 & 1.73 & 20.03 & 17.54 &   20.59 &   17.64 \\
361 & 222.40059 &   8.89156 & 1.61 & 21.62 & 17.90 &    7.61 &   16.55 \\
326 & 222.31325 &   8.87259 & 0.11 & 24.19 & $>19.2$ &    0.88 &    0.00 \\
367 & 222.39220 &   8.88237 & 0.76 & 23.14 & 19.01 &    2.96 &   14.77 \\
 66 & 222.41061 &   9.13401 & 3.71 & 21.81 & 19.10 &    0.80 &    2.04 \\
150 & 222.29356 &   9.05019 & 0.93 & 23.90 & $>19.2$ &    0.77 &    0.00 \\
219 & 222.49631 &   8.99483 & 2.15 & 20.77 & 18.57 &   12.87 &    9.67 \\
209 & 222.39636 &   9.00872 & 0.64 & 23.01 & 18.05 &    3.03 &   22.96 \\
293 & 222.35330 &   8.94158 & 2.81 & 22.60 & 19.14 &    1.16 &    4.92 \\
181 & 222.45479 &   9.03205 & 1.21 & 18.60 & 15.75 &  109.39 &   13.63 \\
239 & 222.30940 &   8.85337 & 0.40 & 23.36 & 17.98 &    1.61 &   23.84 \\
328 & 222.49463 &   8.91316 & 2.70 & 19.73 & 17.36 &   20.44 &    9.25 \\
310 & 222.38542 &   8.87228 & 1.43 & 22.22 & 18.89 &    4.65 &   11.85 \\
175 & 222.32431 &   9.03741 & 1.44 & 23.85 & $>19.2$ &    0.65 &    0.00 \\
360 & 222.33760 &   8.88899 & 1.21 & 18.29 & 15.33 &  220.04 &   18.91 \\
330 & 222.45471 &   8.91179 & 0.65 & 24.62 & $>19.2$ &    0.12 &    0.00 \\
\noalign{\smallskip}\hline
\end{tabular}
\end{minipage}
\end{table*}                                                                   
\newpage
\begin{table*}
\begin{minipage}{0.99\textwidth}

\footnotesize
\begin{tabular}{cccccccc}
\hline\hline\noalign{\smallskip}                                              
ID     &
RA(X)       &    
DEC(X)      &
$\Delta$(X-O) &
R      &
K      &     
LR(R)      &
LR(K)    \\
\noalign{\smallskip}\hline\noalign{\smallskip} 	
186 & 222.39970 &   9.03090 & 0.28 & 23.30 & 18.70 &    1.63 &   15.92 \\
244 & 222.26262 &   8.98473 & 0.84 & 25.22 & $>18.8$&    0.00 &    0.00 \\
178 & 222.27762 &   9.03575 & 1.11 & 19.90 & 18.19 &   26.82 &   16.12 \\
249 & 222.29007 &   8.97873 & 0.83 & 21.33 & $>19.2$ &   10.13 &    0.00 \\
146 & 222.41489 &   9.05666 & 2.31 & 24.98 & 19.11 &    0.00 &    7.23 \\
210 & 222.39806 &   9.01260 & 0.40 & 23.88 & $>19.2$ &    0.86 & 0.00 \\
189 & 222.26312 &   9.02621 & 2.02 & 23.76 & 19.04 &    0.89 &    8.73 \\
250 & 222.43523 &   8.97954 & 1.36 & 23.19 & 18.41 &    2.45 &   14.69 \\
263 & 222.36024 &   8.97151 & 2.47 & 15.07 & 11.76 &  105.64 &   93.10 \\
176 & 222.42137 &   9.03696 & 2.11 & 20.28 & 18.26 &   16.54 &    9.94 \\
251 & 222.29749 &   8.97502 & 2.66 & 20.25 & 16.70 &   11.15 &    8.86 \\
287 & 222.29201 &   8.94793 & 2.22 & 14.61 & 11.86 &   50.25 &  111.02 \\
315 & 222.23370 &   8.92656 & 1.82 & 20.84 & 18.97 &    9.66 &    9.79 \\
242 & 222.35516 &   8.98782 & 1.51 & 21.61 & 17.89 &    7.98 &   17.33 \\
236 & 222.32794 &   8.85342 & 0.89 & 21.74 & 19.33 &    9.98 &   12.88 \\
118 & 222.18840 &   9.08309 & 1.45 & 21.92 & 18.21 &    4.61 &   14.40 \\
 68 & 222.31241 &   9.12831 & 1.75 & 21.03 & 19.04 &   10.03 &   10.17 \\
 94 & 222.43179 &   9.10102 & 2.64 & 22.11 & 17.81 &    2.22 &    8.57 \\
363 & 222.46667 &   8.88736 & 2.22 & 23.33 & $>19.2$ &    0.79 &    0.00 \\
289 & 222.44257 &   8.85969 & 3.16 & 22.18 & 18.36 &    1.41 &    4.33 \\
273 & 222.18602 &   8.96341 & 1.67 & 23.73 & 18.06 &    1.09 &   16.06 \\
301 & 222.32642 &   8.93809 & 1.06 & 20.94 & 17.58 &   13.42 &   23.35 \\
125 & 222.37196 &   9.07913 & 1.85 & 14.90 & 12.56 &  142.06 &   81.20 \\
259 & 222.34471 &   8.97245 & 0.41 & 22.31 & $>19.2$ &    3.70 &    0.00 \\
 77 & 222.34190 &   9.11713 & 0.97 & 23.50 & $>19.2$ &    1.43 &    0.00 \\
139 & 222.39679 &   9.06015 & 0.58 & 22.56 & 18.61 &    3.61 &   18.44 \\
319 & 222.28973 &   8.92027 & 3.30 & 20.67 & 16.98 &    5.02 &    5.00 \\
180 & 222.31947 &   9.03277 & 1.37 & 22.65 & $>19.2$ &    2.86 &    0.00 \\
203 & 222.31352 &   9.01312 & 2.90 & 20.01 & 18.47 &    9.13 &    5.49 \\
 73 & 222.26274 &   9.12213 & 0.39 & 22.20 & $>18.8$ &    6.18 &    0.00 \\
122 & 222.48824 &   9.08229 & 0.90 & 23.18 & 19.00 &    2.86 &   14.26 \\
158 & 222.25020 &   9.04792 & 2.84 & 19.17 & 15.56 &   30.30 &    7.02 \\
282 & 222.22620 &   8.94736 & 3.23 & 21.39 & 18.43 &    2.34 &    4.05 \\
303 & 222.36116 &   8.93674 & 2.67 & 22.23 & 17.90 &    2.17 &    8.37 \\
129 & 222.40411 &   9.07376 & 1.63 & 24.73 & $>19.2$ &    0.04 &    0.00 \\
\hline
\noalign{\smallskip}\hline
\end{tabular}
\end{minipage}
\end{table*}                                                                   
\newpage
\begin{table*}
\begin{minipage}{0.99\textwidth}

\footnotesize
\begin{tabular}{cccccccc}
\hline\hline\noalign{\smallskip}                                              
ID     &
RA(X)       &    
DEC(X)      &
$\Delta$(X-O) &
R      &
K      &     
LR(R)      &
LR(K)    \\
\noalign{\smallskip}\hline\noalign{\smallskip} 	
 67 & 222.40207 &   9.13641 & 1.67 & 22.74 & $>19.2$& 2.50 &    0.00 \\
    &           &           & 3.86 & 21.70 & 18.89 &    1.20 &    1.72 \\ 
138 & 222.26871 &   9.06119 & 1.57 & 22.48 & 18.61 &    2.62 &   13.40 \\
    &           &           & 2.15 & 20.92 & 18.24 &    7.93 &    9.69 \\     
168 & 222.26259 &   9.04336 & 1.77 & 22.44 & 18.78 &    2.37 &   10.06 \\
    &           &           & 1.56 & 21.01 & $>18.8$&11.02 &    0.00 \\ 
103 & 222.27232 &   9.09741 & 3.85 & 20.75 & 17.12 &    2.78 &    2.77 \\
    &           &           & 2.38 & 21.17 & $>18.8$& 1.04 &    0.00 \\
198 & 222.49577 &   9.01765 & 4.04 & 22.90 & 18.49 &    0.28 &    1.67 \\
    &           &           & 1.75 & 23.55 & $>19.2$& 1.04 &    0.00 \\ 
298 & 222.30972 &   8.94036 & 4.37 & 22.28 & 19.15 &    0.36 &    0.82 \\
    &           &           & 1.54 & 23.00 & $>19.2$& 2.26 &    0.00 \\
    &           &           & 3.05 & 23.35 & $>19.2$& 0.41 &    0.00 \\
 61 & 222.33224 &   9.13775 & 4.10 & 23.58 & $>19.2$& 0.13 &    0.00 \\
    &           &           & 4.84 & 23.75 & 18.90 &    0.05 &    0.48 \\
237 & 222.33333 &   8.98678 & 1.87 & 24.64 & 19.11 &    0.07 &    9.53 \\
    &           &           & 3.26 & 21.96 & 18.94 &    1.28 &    3.26 \\
\hline
362 & 222.43140 &   8.88969 & 5.71 & 23.13 & $>19.2$& 0.02 &    0.00\\        
 23 & 222.37933 &   9.17607 & 3.95 & 23.31 & $>18.8$& 0.07 &    0.00 \\
    &           &           & 3.47 & 23.17 & $>18.8$& 0.00 &    0.00 \\ 
357 & 222.42833 &   8.88944 & 4.69 & 23.79 & $>19.2$& 0.06 &    0.00 \\
    &           &           & 4.87 & 24.89 & $>19.2$& 0.00 &    0.00 \\
157 & 222.23581 &   9.04881 & 4.31 & 23.55 & $>18.8$& 0.10 &    0.00 \\
    &           &           & 5.23 & 20.32 & 16.47 &    0.42 &    0.19 \\     
107 & 222.23543 &   9.09159 & 1.55 & 25.76 & $>18.8$& 0.00 &    0.00 \\
    &           &           & 2.39 & 25.24 & $>18.8$& 0.00 &    0.00 \\
 28 & 222.27785 &   9.16305 & 3.84 & 24.14 & $>18.8$& 0.10 &    0.00 \\
    &           &           & 0.86 & 25.38 & $>18.8$& 0.00 &    0.00 \\
    &           &           & 4.35 & 24.96 & $>18.8$& 0.00 &    0.00 \\
\hline
     269 & 222.24790 &   8.96716 & --   & --     & --     & --     & -- \\ 
     255 & 222.35733 &   8.97293 & --  & -- &  -- & -- & -- \\ 
     288 & 222.33347 &   8.94492 & -- &  -- & -- & -- & -- \\ 
     152 & 222.36362 &   9.05004 & -- &  -- & -- & -- & -- \\ 
      154 & 222.39186 &   9.05019 & -- & -- & -- & -- & -- \\ 
     207 & 222.48463 &   9.01196 & --  & -- &  -- & -- & -- \\ 
%
     179 & 222.17659 &   9.03417 & --  & -- &  -- & -- & -- \\ 
     344 & 222.23007 &   8.90152 & --  & -- &  -- & -- & -- \\ 
     241 & 222.21679 &   8.98637 & --  & -- & -- & -- & -- \\ 
\noalign{\smallskip}\hline
\end{tabular}

\vspace{0.5cm}
Notes: ID = X--ray source identifications; RA(X), DEC(X) = X--ray
coordinates; R,K = R--band and K--band magnitudes;
LR(R), LR(K)= Likelihood ratio in the R and K bands. 
\end{minipage}
\end{table*}                                                       

\begin{table*}
       \caption{Hard X--ray detected EROs}
       \label{tab_eros_1}
\begin{minipage}{0.99\textwidth}
\footnotesize
\begin{tabular}{rccccccccccc}
\hline\hline\noalign{\smallskip}                                              
ID     &
RA(X)       &    
DEC(X)      &
$\Delta$(X-O) &
R (2$''$)      &
K (2$''$)      &     
R-K    &
F$_{2-10 keV}$ & 
HR      &
log(X/O)  &
log(X/K)  & 
 LR(K) \\
 & (J2000) & (J2000) & $''$ &  &  & & \cgs & & & & \\
\noalign{\smallskip}\hline\noalign{\smallskip} 	
\multicolumn{12}{c}{5$\sigma$ ERO sample} \\ 
148 & 222.40681 &   9.05575 & 0.37 & 25.24 & 18.76 & 6.48 & 
	2.9e-14 & -0.44 & 2.06   &   0.97  & 15.78 \\
195 & 222.23720 &   9.02275 & 1.73 & 24.11 & 18.32 & 5.79 &  
	2.5e-14  &  0.00 & 1.54  & 0.73  & 15.58 \\
209 & 222.39636 &   9.00872 & 0.64 & 23.27 & 18.08 & 5.19 & 
	9.5e-15  &  -0.54 & 0.78  &    0.21 & 22.96 \\ 
293 & 222.35330 &   8.94158 & 2.81 & 24.26 & 19.13 & 5.13 & 
        9.1e-15  & $>0.76$ & 1.16  & 0.61  & 4.92 \\
239 & 222.30940 &   8.85337 & 0.40 & 23.96 & 18.23 & 5.73 & 
        8.9e-15   & -0.19 & 1.03  & 0.24 & 23.84 \\ 
146 & 222.41489 &   9.05666 & 2.31 & 25.63 & 19.10 & 6.53 &  
        6.5e-15  &  0.01 &  1.56  & 0.45  & 7.23 \\
189 & 222.26312 &   9.02621 & 2.02 & 24.40 & 19.04 & 5.36 & 
        5.8e-15  & $>0.32$ & 1.02  & 0.38  & 8.73 \\
250 & 222.43523 &   8.97954 & 1.36 & 23.50 & 18.47 & 5.03 & 
        5.8e-15  & $>0.42$ & 0.66  & 0.15  & 14.69 \\
%
%
\hline
\multicolumn{12}{c}{ 3$\sigma$ ERO sample} \\ 
     175 & 222.32430 &   9.03741 & 1.44 & 24.39 &  19.35 & 5.04  &
      7.9e-15  & 0.12 & 1.24  & 0.43 & -- \\
%
%
%
\noalign{\smallskip}\hline
\end{tabular}
\end{minipage}
\end{table*}                                                                         
%
%

\end{document}